\documentclass[iopart,reprint,preprintnumbers,amsmath,amssymb,showkeys,showpacs,nofootinbib]{revtex4-1}
\usepackage{graphicx}% Include figure files
\usepackage{dcolumn}% Align table columns on decimal point
\usepackage{bm}% bold math
\usepackage{rotating}% bold math
\usepackage{xcolor}% math symbols in color
\usepackage{verbatim}
\usepackage{stmaryrd}
\usepackage{appendix}
\usepackage{cancel}
\usepackage{array}
\DeclareMathOperator{\erf}{erf}

\begin{document}

\title{Spatially-dependent modeling and simulation of runaway electron
  mitigation in DIII-D}

\author{M.~T.~Beidler}\thanks{corresponding author:
  beidlermt@ornl.gov} \affiliation{Oak Ridge National Laboratory, Oak
  Ridge, Tennessee 37831-8071, USA}

\author{D.~del-Castillo-Negrete} \affiliation{Oak Ridge
  National Laboratory, Oak Ridge, Tennessee 37831-8071, USA}
\author{L.~R.~Baylor} \affiliation{Oak Ridge
  National Laboratory, Oak Ridge, Tennessee 37831-8071, USA}
\author{D.~Shiraki} \affiliation{Oak Ridge
  National Laboratory, Oak Ridge, Tennessee 37831-8071, USA}
\author{D.~A.~Spong} \affiliation{Oak Ridge
  National Laboratory, Oak Ridge, Tennessee 37831-8071, USA}

%\preprint{Draft: \today}

%\date{\today}

\begin{abstract}
New simulations with the Kinetic Orbit Runaway electron (RE) Code KORC
show RE deconfinement losses to the wall during plasma scrape off are
the primary current dissipation mechanism in DIII-D experiments with
high-Z impurity injection, and not collisional slowing down. The
majority of simulations also exhibit an increase in the RE beam energy
due to acceleration by the induced toroidal electric field, even while
the RE beam current is decreasing. In this study, KORC integrates RE
orbits using the relativistic guiding center equations of motion, and
incorporates time-sequenced, experimental reconstructions of the
magnetic and electric fields and line integrated electron density to
construct spatiotemporal models of electron and partially-ionized
impurity transport in the companion plasma. Comparisons of
experimental current evolution and KORC results demonstrate the
importance of including Coulomb collisions with partially-ionized
impurity physics, initial RE energy, pitch angle, and spatial
distributions, and spatiotemporal electron and partially-ionized
impurity transport. This research provides an initial quantification
of the efficacy of RE mitigation via injected impurities, and
identification of the critical role played by loss of confinement due
to plasma scrape off on the inner wall as compared to the relatively
slow collisional damping.
\end{abstract}

\maketitle

\section{Introduction}

If not avoided or dissipated, runaway electrons (REs) can seriously
damage ITER's plasma-facing components \cite{Hender07,Boozer15}. If
avoidance fails, shattered pellet injection (SPI) is the leading
candidate to dissipate REs in ITER
\cite{Commaux10,Lehnen15,Hollmann15b,Baylor19}, and although
significant progress has been made, there is a pressing need for
modeling and simulation studies to assess the efficacy of SPI and to
optimize different dissipation strategies, such as massive gas
injection (MGI). From the theory and simulation perspective, there is
a need to develop and validate realistic models of the interaction of
REs with partially-ionized impurities, such as those proposed in
Refs.~\cite{Hesslow17,Hesslow18}.

\begin{figure}
\centering
\noindent\includegraphics[width=3.4in]{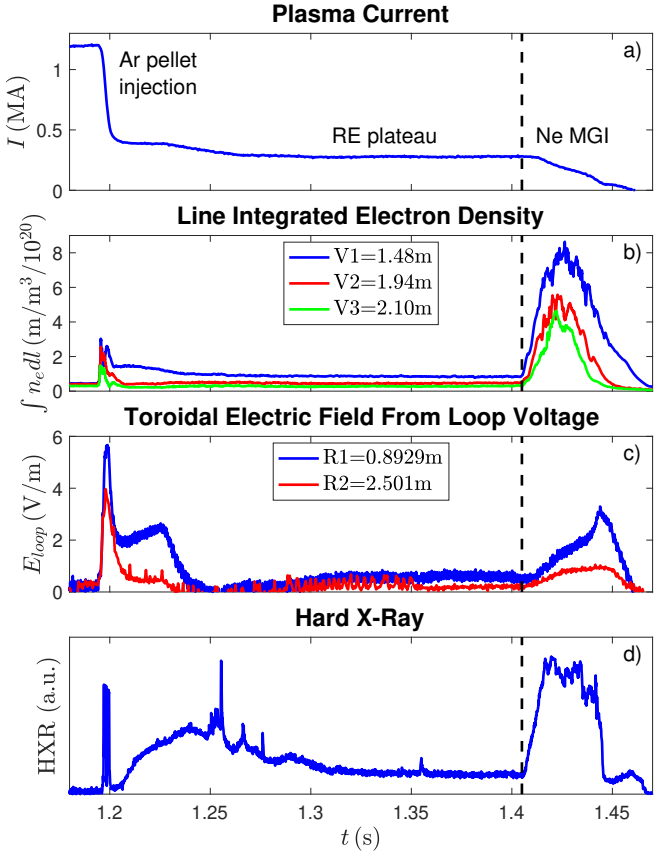}
\caption[]{\label{fig-d3draw}DIII-D discharge \#164409 with Ne MGI
  into the post-disruption, RE plateau, dissipating the RE current in
  panel a). The impurity injection results in greatly increased
  densities at time $1.405\,{\rm ms}$ (marked by vertical, dashed
  line) as seen by interferometer chord data in b). Changes in RE
  current after Ne MGI drive a toroidal loop voltage consistent with a
  toroidal electric field in panel c). REs interacting with the
  companion plasma and limiters produce hard x-rays in panel d). }
\end{figure}

Recent experiments at DIII-D indicate that SPI and MGI perform
similarly \cite{Shiraki18}. This work will focus on DIII-D discharge
\#164409, which has been reported extensively in
Ref. \cite{Shiraki18}. The evolution of characteristic parameters are
plotted in Fig.~\ref{fig-d3draw}. At approximately $1.2\,{\rm s}$, a
small Ar pellet is injected that triggers the observed current quench
in Fig.~\ref{fig-d3draw}a. A RE beam is generated and position
controlled until a secondary injection of Ne gas at approximately
$1.4\,{\rm s}$ (marked by the vertical, dashed line) causes the RE
beam to dissipate. Figure \ref{fig-d3draw}b shows three vertical,
interferometer chords, capturing the evolution of line integrated
electron density. Figure \ref{fig-d3draw}c shows the inferred toroidal
electric field from the loop voltage measured at the high field side
(HFS) and low field side (LFS).  The bremsstrahlung hard x-ray (HXR)
signals in Fig.\ref{fig-d3draw}d are measured by a bismuth-germanate
based scintillator located at the bottom of DIII-D and mostly
sensitive to $1-10,{\rm MeV}$ photons \cite{James10}.  The RE plateau
phase is significantly different from the pre-disruption, in that the
majority of the current is carried by REs and not by the cold
($T_e<2\,{\rm eV}$) ``companion'', or background, plasma
\cite{Hollmann13,Hollmann11}. For this reason, the dynamics during
this phase can be well studied using a particle tracking code, and in
this work we employ the Kinetic Orbit Runaway electrons Code (KORC)
\cite{Carbajal17a}. The data in Fig.~\ref{fig-d3draw} will be both
used as inputs to KORC calculations as described in
Sec.~\ref{sec::model}, and compared to outputs from KORC calculations
in Sec.~\ref{sec::results}.

There have been significant previous efforts to model RE dissipation
by impurity
injection. Refs.~\cite{Aleynikov14,Konovalov16,MartinSolis17} use 1D
transport codes with Fokker-Planck models for REs having physics of
bound electrons and partially-ionized impurities. Ref.~\cite{Reux15}
uses ESTAR modeling of the Bethe stopping power of an idealized beam
of REs in JET-ILW. Refs.~\cite{Spong18,dCN18} use guiding center test
particle modeling using KORCGC, a predecessor of KORC, with 2D
axisymmetric fields and Monte Carlo collision operators having physics
of bound electrons and partially-ionized impurities with constant
density profiles. 

Several studies using guiding center test particle modeling have also
been employed to study RE generation and confinement during the
thermal quench of a disruption. Refs.~\cite{Papp11a,Papp11b} use the
ANTS code to evolve particle orbits in 3D
fields. Refs.~\cite{Sommariva17,Sommariva18} use the RE orbit module
in JOREK to calculate RE confinement and hot tail and Dreicer
generation during a simulated thermal quench. Ref.~\cite{Izzo11} uses
the RE orbit module in NIMROD to calculate RE confinement during
impurity injection induced thermal quench and Ref.~\cite{Jiang19} uses
the same module to calculate RE confinement with a pre-seeded
large-scale island structure. Ref.\cite{McDevitt19} uses a RE orbit
module with Monte Carlo collision operators having physics of bound
electrons and partially-ionized impurities to calculate the post
thermal quench spatial distribution of REs. Ref.\cite{Liu19} uses a RE
orbit module in MARS-F to calculate RE loss by magnetohydrodynamic
(MHD) instabilities. The confinement of REs in stochastic magnetic
fields characteristic of the thermal quench phase was studied in
Ref.~\cite{Carbajal20}.

Additional previous studies have developed tools for modeling the
transport of injected impurities. Ref.~\cite{Whyte03} developed KPRAD
to study the effect of MGI on disruptions.  Ref.~\cite{Nardon16}
developed the 1D radial fluid code IMAGINE to model MGI with
comparisons to JET interferometer diagnostics. Ref.~\cite{Hollmann19}
recently developed a 1D diffusion model for impurity profiles and
evolution.

The present study builds on previous research by incorporating
experimentally-reconstructed, time-dependent magnetic and electric
fields and employing models for the spatial-dependence of the
injected impurity and electron density. Additionally, the flexible
KORC framework is used to simulate RE dissipation with different
models for bound electrons and partially-ionized impurity physics.
We note that the physics of large-angle collisions comprising a source
of secondary RE generation is not included in the present modeling,
but implementation is underway for future studies.  The code KORC has
been extended to serve as a general framework for simulating RE
physics, including validation and verification of the theoretical
models needed to understand RE dissipation by impurity injection.

A major theme and contribution of the present paper is the assessment
of the effectiveness of RE dissipation given the competing time scales
of the RE loss of confinement (due to the displacement of the flux
surfaces and the eventual loss of magnetic confinement) and RE energy
dissipation.  The remainder of this paper is organized as follows.  In
Sec.~\ref{sec::model} we introduce the extensions to KORC permitting
the study of RE dissipation by impurity injection, including
relativistic guiding center equations of motion, incorporating
experimentally-reconstructed magnetic and toroidal electric fields,
models for a spatiotemporal density profile, Monte Carlo linearized,
Coulomb collision operator with models of bound electron effects, and
synchrotron and bremsstrahlung radiation models.  In
Sec.~\ref{sec::init} we discuss the initialization of the RE
distributions used in KORC simulations. In Sec.~\ref{sec::results} we
present results of KORC simulations and comparisons to DIII-D
experiment \#164409. Lastly, in Sec.~\ref{sec::conclusion} we provide
concluding remarks.

\section{Physics Model}
\label{sec::model}

\subsection{Relativistic Guiding-Center Equations}

\begin{figure*}
\centering
\noindent\includegraphics[width=5.4in]{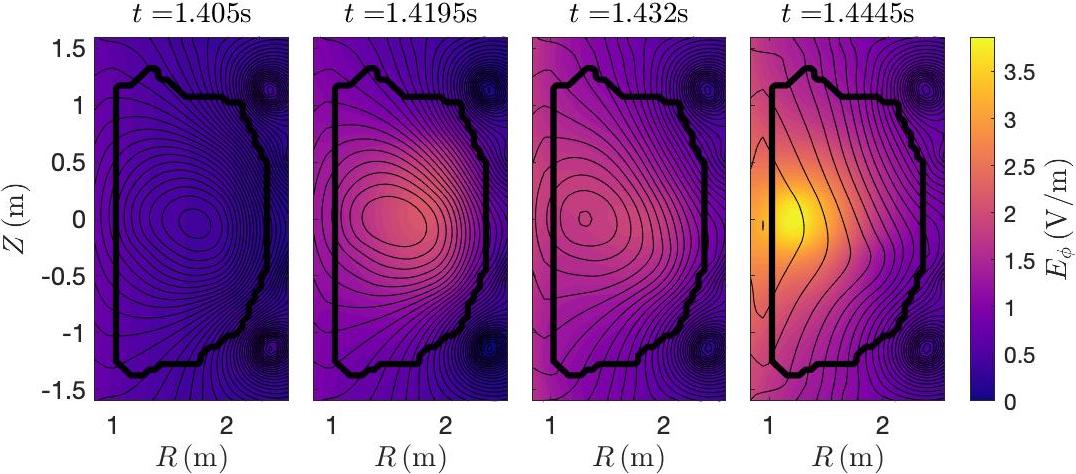}
\caption[]{\label{fig-psiE}Snapshots of the time-evolving, poloidal
  flux contours (thin, black contours) from DIII-D discharge
  \#164409 as calculated with JFIT. The overlaid thick, black contour
  shows the approximate first wall of DIII-D. The colormap indicates
  the toroidal electric field as calculated from the time-derivative
  of the poloidal flux contours.}
\end{figure*}

The original development of KORC in Ref.~\cite{Carbajal17a} stressed
the importance of full orbit (FO) effects in RE
calculations. Inclusion of FO effects enabled accurate calculation of
synchrotron emission \cite{Carbajal17b} and comparison with
experimental observations \cite{dCN18}. For the present study,
retaining FO effects yields calculations that are prohibitively
numerically expensive. To make modeling RE orbits for the duration of
the RE dissipation phase numerically feasible, we employ the
relativistic guiding center (RGC) model from
Refs.~\cite{Tao07,Cary09}
\begin{subequations}
\begin{align}
  \frac{d{\bf X}}{dt}&=\frac{1}{{\bf b}\cdot{\bf B}^*}\Bigg(e{\bf
    E}\times{\bf b}-p_\parallel\frac{\partial{\bf b}}{\partial
    t}\times{\bf b} \nonumber \\ &\hspace{0.7in}+\frac{m\mu{\bf
      b}\times\nabla B+p_\parallel{\bf B}^*}{m\gamma_{\rm
      gc}}\Bigg) \label{eq-gcx} \\ \frac{dp_\parallel}{dt}&=\frac{{\bf
      B}^*}{{\bf b}\cdot{\bf B}^*}\cdot\left(e{\bf
    E}-p_\parallel\frac{\partial{\bf b}}{\partial t}-\frac{\mu\nabla
    B}{\gamma_{\rm gc}}\right), \label{eq-gcp}
\end{align}
\end{subequations}
where ${\bf X}\in\mathbb{R}^3$ denotes the spatial location of the GC
in cylindrical $(R,\phi,Z)$ coordinates and $p_\parallel\in\mathbb{R}$
denotes the component of the relativistic momentum along the magnetic
field, $p_\parallel\equiv \gamma m ({\bf V}\cdot{\bf b})=\gamma m
V\cos\eta$, with ${\bf V}=d{\bf X}/dt$ the velocity of the GC, ${\bf
  b}={\bf B}/B$ the unit magnetic field vector, $\eta$ the pitch
angle, $m$ the particle mass, $e$ the particle charge,
$\gamma=\left[1-(V/c)^2\right]^{-1/2}$, and the magnitude of a vector
${\bf A}$ given by $A=\sqrt{{\bf A}\cdot{\bf A}}$. The magnetic moment
is defined as
\begin{equation}
  \mu=\frac{|{\bf p}-p_\parallel{\bf b}|^2}{2mB}=\frac{p_\perp^2}{2mB},
\end{equation}
and is assumed constant in the absence of collisions and radiation,
with $p_\perp=\gamma m V\sin\eta$. The ``effective'' magnetic field
is defined as
\begin{equation}
  {\bf B}^*=q{\bf B}+p_\parallel\nabla\times{\bf b},
\end{equation}
and the GC relativistic factor is defined as
\begin{equation}
  \gamma_{\rm gc}=\sqrt{1+\left(\frac{p_\parallel}{mc}\right)^2+
    \frac{2\mu B}{mc^2}}.
\end{equation}
Note that the RGC model is also used in Ref.~\cite{Carbajal20}, where
their Fig.~15 compares FO and RGC orbits.  Planned future work will
utilize output from simulations using the RGC model to initialize
shorter duration calculations with the FO model for evaluating
synchrotron emission to be used in comparisons with experimental
observations.

In the present work, we assume the static magnetic field limit of the
RGC equations, by ignoring the second terms in the parenthesis of both
Eqs.~(\ref{eq-gcx}) and (\ref{eq-gcp}). This can be shown to be valid
by scaling the right-hand-side (RHS) velocity terms for the spatial
location
\begin{subequations}
\begin{align}
  &e{\bf E}\times{\bf b}& &\sim& &e& &\sim&
  &10^{-19}& \label{eq-scaleE}\\ &p_\parallel\frac{\partial{\bf
      b}}{\partial t}\times{\bf b}& &\sim& &\gamma m_ec& &\sim&
  &10^{-21}& \label{ee-scaledt}\\ &\frac{\mu}{\gamma} {\bf
    b}\times\nabla B& &\sim& &\gamma m_ec^2\eta^2& &\sim&
  &10^{-14}& \label{eq-scalegrB}\\ &\frac{p_\parallel{\bf
      B}^*}{m_e\gamma}& &\sim& &ec& &\sim& &10^{-11}&
\label{eq-scalepar}
\end{align}
\end{subequations}
where we assume a RE has kinetic energy of $10\,{\rm MeV}$ and pitch
angle of $10^\circ$, electric field, magnetic field, and spatial
length scale are of order unity, and Faraday's law is used to scale
$\partial{\bf b}/\partial t$.  The term coming from the covariant
relativistic correction proportional to the time-derivative of the
magnetic unit vector is the smallest contribution, and many orders of
magnitude smaller than the leading terms. More precise calculations
(not shown) also indicate the applicability of the static magnetic
field limit used in the equations of motion in this work. This limit
is taken because calculating the time-derivative of the magnetic field
would double the number of interpolations needed, leading to
additional computational costs.

The model equations are integrated employing the Cash-Karp 5th order
Runge-Kutta method \cite{Cash90}. For an axisymmetric magnetic
configuration, in the absence of an electric-field, collisions, or
radiation, energy and the canonical toroidal angular momentum are
conserved. These conserved quantities are used to test the accuracy of
orbit calculations in an axisymmetric magnetic configuration without
electric field, collisions, or radiation. Note that the calculations
in Sec.~\ref{sec::results} do not have these restrictions and energy
and canonical toroidal angular momentum are not conserved. We find
that the accuracy of calculations are dependent on resolving the
magnitude of magnetic curvature in the configuration, dominated by the
motion parallel to the magnetic field, consistent with
Eq.~(\ref{eq-scalepar}). Based on the results of a convergence study
of time step (not shown), we use $dt=eB_0/\gamma
m_e=3.4321\times10^{-10}\,({\rm s})$ in all simulations, calculated
for a RE with kinetic energy of $10\,{\rm MeV}$.  

\subsection{Plasma Model}
\label{sec-plasmamodel}

The magnetic field components in the poloidal-plane are calculated
from the poloidal flux function $\psi_p$ as determined from JFIT
reconstructions \cite{Humphreys99} of DIII-D discharges. The JFIT
reconstruction differs from the more often used EFIT reconstruction
\cite{EFIT}, in that it doesn't find an Grad-Shafranov equilibrium,
instead calculating poloidal flux from a best-fit toroidal current
distribution composed of distributed-current (plasma) elements as well
as passive structure currents (such as sections of the vacuum vessel),
as constrained by magnetic diagnostics and measured equilibrium coil
currents. In the RE plateau, due to the low temperature and density of
the companion plasma, the neutral beams needed for Motional Stark
Effect \cite{Levinton89} magnetic diagnostic are not available, due to
the risk of shine through and damage to the first wall. Thus, there
are no internal magnetic diagnostics available for constraining the
JFIT equilibrium used in this study. However, because the constraints
provided by magnetic diagnostics near the edge of the plasma ensure a
reliable reconstruction of the last closed flux surface, and the
primary concern of this work is to assess the comparative loss of
confinement and collisional dissipation of REs, JFIT reconstructions
are suitable for this study.

\begin{figure}
\centering
\noindent\includegraphics[width=3.4in]{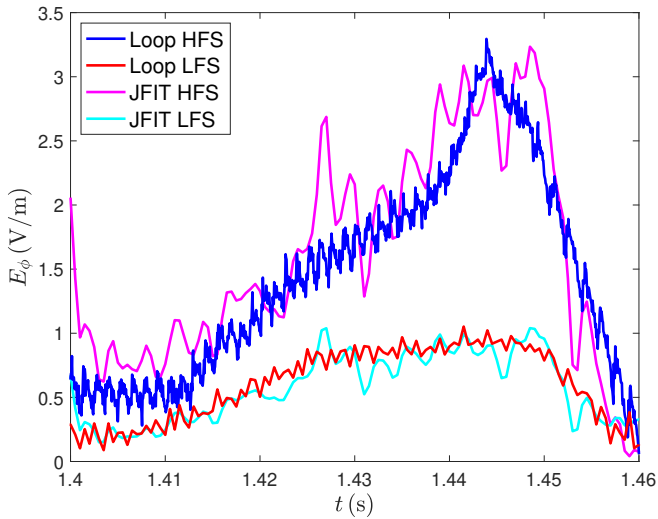}
\caption[]{\label{fig-Eevocompare} Measurements of the toroidal
  electric field calculated via the toroidal loop voltage for DIII-D
  experiment \#164409 at the HFS $R=0.8929\,({\rm
    m})$ (dark, blue trace) and the LFS
  $R=2.5010\,({\rm m})$ (red trace). Point measurements of the
  toroidal electric field calculated via time-derivatives of the JFIT
  poloidal flux function are shown at the HFS (violet trace) and LFS
  (cyan trace).}
\end{figure}

\begin{figure*}
\centering
\noindent\includegraphics[width=6.75in]{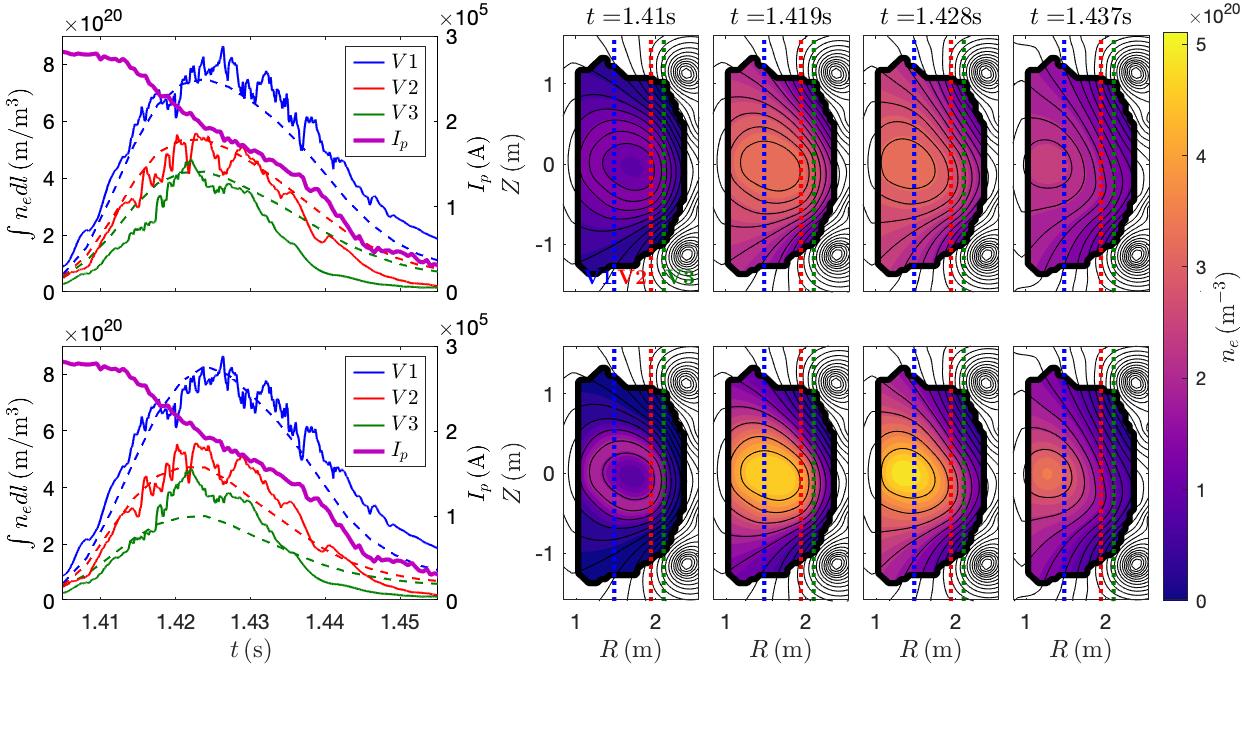}
\caption[]{\label{fig-denevocompare} Left plots show experimental
  current and line integrated density measurements in solid traces for
  DIII-D discharge \#164409. Dashed traces indicate synthetic
  interferometer diagnostic results using the spatiotemporal density
  profile from Eqs.~(\ref{eq-ne})-(\ref{eq-sige}), where the top plot
  uses a broad and diffuse profile with parameters in the top row of
  Table \ref{tab-denevo} and the bottom plot uses a narrow and dense
  profile with parameters from the bottom row. Right plots show
  spatiotemporal density profiles at different time slices, where the
  vertical dashed lines indicate the location of the interferometer
  diagnostics on DIII-D. Note that the different spatiotemporal
  profiles record similar synthetic interferometer signals.}
\end{figure*}

KORC uses JFIT-computed $\psi_p$ on a $(NR\times NZ)=(33\times 65)$
grid, which is then interpolated using the \texttt{PSPLINE} cubic
spline interpolation routines \cite{PSPLINE}. In the cylindrical
coordinate frame $(R,\phi,Z)$ we adopt the convention ${\bf
  B}_p={\boldsymbol \nabla}\phi\times{\boldsymbol \nabla}\psi_p$ for
the poloidal field components yielding $B_R=(1/R)\partial
\psi_p/\partial Z$, $B_Z=-(1/R)\partial \psi_p/\partial R$, where the
gradients are computed using the spline representation in
\texttt{PSPLINE}. In the discharges of interest the current is
directed in the positive $\phi$ direction, giving $\psi_p$ a local
minimum at the magnetic axis. The toroidal magnetic field component is
calculated assuming $B_\phi=-R_0B_0/R$, directed in the counter-$I_p$
direction, where $R_0=1.682{\rm m}$ and $B_0=2.141{\rm T}$ are the
major radial location and toroidal field near the magnetic axis for
DIII-D discharge \#164409. This model for the toroidal magnetic field
does not include the self-consistent correction due to poloidal RE
motion, however simple calculations (not shown) indicate that this correction
should be on the order of $0.1\%$ In addition to calculating the
magnetic field components by taking the first derivatives using the
\texttt{PSPLINE} routines, we also calculate the second order
derivatives from the spline representation of $\psi_p$ to compute the
${\boldsymbol \nabla}B$ and ${\boldsymbol \nabla}\times{\bf b}$
``auxiliary'' fields needed to integrate
Eqs.~(\ref{eq-gcx}-\ref{eq-gcp}).

In this work, we use time-sequenced JFIT reconstructions of $\psi_p$
that are spaced $0.5\,{\rm ms}$ apart. We preprocess this data with
\texttt{MATLAB}, first smoothing $\psi_p$ over $5\,{\rm ms}$
intervals, and then taking the time-derivative using a central
difference method to calculate $E_\phi$ according to $E_\phi=(1/2\pi
R)\partial\psi_p/\partial t$. Snapshots of the smoothed $\psi_p$
contours are shown in Fig.~\ref{fig-psiE}, by the thin contours, where
the thick contour indicates the DIII-D first wall.  The experimentally
inferred magnetic fields are by construction axisymmetric and thus
they are not stochastic. Although exploring the potential role of
magnetic field stochasticity is outside the scope of the present
paper, we believe that in this case this effect is much less important
than the fast deconfinement due to the scrap off. It is also important
to keep in mind that is some regimes the deconfinement of RE due to
magnetic stochasticity is less that commonly thought, see for example
Ref.~\cite{Carbajal20}. As the current decays, Fig.~\ref{fig-psiE}
indicates that flux surfaces advect toward the inner wall limiter. At
present, there are two possibilities for this advection:
ineffectiveness of the position controller at low currents
\cite{Eidietis12}, or evolution of canonical toroidal angular momentum
as REs decelerate \cite{Hu16}.  It is outside of the scope of this
work to determine the cause of this advection, but rather analyze the
consequences for RE confinement and collisional dissipation. The
calculated toroidal electric field is shown in Fig.~\ref{fig-psiE} by
the colormap. To check the robustness of the method for calculating
$E_\phi$, we compare values calculated from the JFIT reconstructions
with data from loop voltages measurements from the HFS and LFS of
DIII-D in Fig.~\ref{fig-Eevocompare}, which indicates good agreement.
At the beginning of each simulation and $5\,{\rm ms}$ intervals
thereafter, \texttt{PSPLINE} recalculates the interpolants for
$\psi_p$ and $E_\phi$ for the next time interval.

The electron temperature of the plasma is assumed temporally and
spatially constant at $T_e=1.5\,{\rm eV}$ as seen from Thomson
scattering measurements taken before impurity injection (not shown),
which is also consistent with fits to line emission brightness in
Ref.~\cite{Hollmann13}. The effective impurity nuclear charge is also
assumed temporally and spatially constant at $Z_{\rm eff}=1$. We will
comment how these assumptions will affect modeling in
Secs.~\ref{sec::reltrans},\ref{sec::bounde}. The electron density
takes the form of a temporally and spatially evolving, constricting
``ring'' profile. This profile is given by that of a Gaussian centered
at a chosen flux surface with a time-dependent magnitude and width
parameterized as
\begin{subequations}
\begin{align}
  n_e(\psi_p,t)&=\frac{n_r-n_0}{2}\left[\tanh\left(\frac{t-\tau_{\rm
        in}}{\tau_{\rm in}}\right) -\tanh\left(\frac{t-t_{\rm
        delay}}{\tau_{out}}\right)\right] \nonumber \\ & \times
  \exp\left[-\frac{\left(\sqrt{\psi_p}-\sqrt{\psi_{p,0}}\right)^2}{2\sigma_{\psi_p}^2(t)}\right] \label{eq-ne}
  \\ \sigma_{\psi_p}(t)&=\lambda_n\erf\left(\frac{t}{\tau_{\rm
      n}}\right) \label{eq-sige}
\end{align}
\end{subequations}
where $n_r$ is the maximum density of the profile, $n_0$ is the
background density, $\tau_{\rm in}$, $\tau_{\rm out}$, and $t_{\rm
  delay}$ parameterize the time scale over which the density
increases, decreases, and remains constant, respectively, $\psi_{p,0}$
the poloidal flux surface where the Gaussian is centered, $\lambda_n$
is the maximum width of the Gaussian, and $\tau_n$ is the time scale
over which the Gaussian width increases.

\begin{table*}[]
  \centering
  \begin{tabular} {c|c|c|c|c|c|c|c}
    $n_r\,({\rm m^{-3}})$ & $n_0\,({\rm m^{-3}})$ & $\tau_{\rm
      in}\,({\rm s})$ & $\tau_{\rm out}\,({\rm s})$ & $t_{\rm
      delay}\,({\rm s})$ & $\psi_{p,0}$ & $\lambda_n $ &
    $\tau_{n}\,({\rm s})$\\ \hline $4\times10^{20}$ &
    $2.5\times10^{19}$ & $7.5\times10^{-3}$ & $1.25\times10^{-2}$ &
    $4\times10^{-2}$ & $0.8$ & $0.255$ & $1.5\times10^{-2}$
    \\ \hline $7\times10^{20}$ & $2.5\times10^{19}$ & $10^{-2}$ &
    $1.25\times10^{-2}$ & $5\times10^{-2}$ & $0.775$ & $0.15$ &
    $1.75\times10^{-2}$
  \end{tabular}
  \caption[]{\label{tab-denevo}Spatiotemporal density profile fitting
    parameters for modeling DIII-D discharge \#164409. Top row
    corresponds the a broad and diffuse profile while the bottom row
    corresponds to a narrow and dense profile.}
\end{table*}

Physically, as impurities are injected via SPI, they first encounter a
cold companion plasma, with little ionization of the impurities. As
the impurities reach the RE beam, the impurities are rapidly ionized
and the resulting electrons move along poloidal flux surfaces. This
ring structure can be seen experimentally by visible cameras (not
shown). Due to the low temperatures of the companion plasma, the
electrons and impurities diffuse across magnetic surfaces
\cite{Hollmann13}. Lastly, as the current decreases, the electrons and
impurities are deconfined due to the motion of the flux surfaces.

This work assumes that all impurity charge states considered have the
same spatiotemporal profile as $n_e$, with only changes in their ratio
$n_j/n_e$. Presently we assume the companion plasma contains two
partially-ionized impurity charge states, with
$n_{Ne^+1}/n_{Ne^+2}=2$. This choice of impurity composition is
roughly in line with results presented in Ref.~\cite{Hollmann19}.
While the companion plasma contains a significant D+ population in the pre-MGI state, Ref.~\cite{Hollmann19} infers that D+ plays an insignificant role in the post-MGI state studied in this work.”

We
will further discuss the calculation of impurity charge state
densities and the affect of the assumed charge state ratio in
Sec.~\ref{sec::bounde}.

In the following simulations, two spatiotemporal profiles are used, a
``broad and diffuse'' profile, seen in the top row of plots in
Fig.~\ref{fig-denevocompare} with parameters in the top row of Table
\ref{tab-denevo}, and a ``narrow and dense'' profile seen in the
bottom row of plots in Fig.~\ref{fig-denevocompare} with parameters in
the bottom row of Table \ref{tab-denevo}. The left column of plots in
Fig.~\ref{fig-denevocompare} compare line-integrated electron density
interferometer diagnostic data from DIII-D discharge \#164409 with
synthetic line-integrated density diagnostic applied to the model
density shown in the right plots of Fig.~\ref{fig-denevocompare}. The
two spatiotemporal density profiles were chosen to test the
sensitivity of the model used given the line-integrated measurements.

\subsection{Coulomb Collisions}
\label{sec-collisions}

\subsubsection{Monte Carlo Operator}

In flux-conserving form, the Fokker-Planck partial differential
equation (PDE) with a Coulomb collision operator in
azimuthally-symmetric spherical momentum space ({\it e.g.}
\cite{Trubnikov65})
\begin{align}
  \mathcal{C}(f)=&\frac{1}{p^2}\frac{\partial}{\partial
    p}\left[p^2\left(C_A\frac{\partial f}{\partial p}+C_F
    f\right)\right] \nonumber \\&+
  \frac{C_B}{p^2}\left[\frac{1}{\sin\eta}\frac{\partial}{\partial\eta}
    \left(\sin\eta\frac{\partial f}{\partial\eta}\right)\right]
\end{align}
can be written \cite{Gardiner04} as two
equivalent stochastic differential equations (SDEs) for the
phase-space momentum, given by
\begin{subequations}
\begin{align}
  dp&=\Bigg\{-C_F(p)+\frac{1}{p^2}\frac{\partial}{\partial
    p}\left[p^2C_A(p)\right]\Bigg\}
  dt\nonumber \\ &+\sqrt{2C_A(p)}\,dW_p, \label{eq-dp}\\ d\eta&=\frac{C_B(p)}{p^2}\cot\eta
  dt+\frac{\sqrt{2C_B(p)}}{p}\,dW_\eta, \label{eq-deta}
\end{align}
\end{subequations}
where $C_F$, $C_A$, and $C_B$ are linearized transport coefficients for
collisional friction (slowing down), parallel diffusion, and pitch
angle scattering (deflection), respectively, and $dW$ is a zero mean,
unit standard deviation Weiner process satisfying
\begin{equation}
  <dW>=0, \hspace{.5in} <(dW)^2>=dt.
\end{equation}
Here, we use uniformly-distributed random numbers, which behave better
than normally-distributed random numbers at low energies, due to their
tighter bounds. Applying It\^{o}'s lemma \cite{Gardiner04} by letting
$\xi=\cos\eta$ yields
\begin{align}
  d\xi&=-2\xi\frac{C_B(p)}{p^2}dt-\frac{\sqrt{2C_B(p)}}{p}
  \sqrt{1-\xi^2}\,dW_{\xi}. \label{eq-dxi}
\end{align}

The Coulomb collision operator SDEs are subcycled independently of the
RGC equations of motion using an operator splitting method. The time
step of the collision operator is set as the $1/20$ of the shortest
inverse collision frequency (to be defined in the following
subsections). While this temporal resolution accurately captures the
damping of relativistic particles, with relatively long inverse
collision frequencies, once particles thermalize this temporal
resolution is not sufficient, consistent with the divergence of
$d\eta$ as $p\rightarrow0$. Thus, for this study, when a particle's
momentum satisfies $p<m_ec$, the particle is flagged as thermalized
and is not tracked anymore. Future studies, especially for the
generation of runaways from a thermal plasma following the thermal
quench, will require a significant increase in the collision substep
cadence. Additionally, we again mention that the physics of
large-angle collisions comprising a source of secondary RE generation
is not included in the present modeling, and its implementation is
expected to increase the collision subcycling frequency substantially.

\subsubsection{Relativistic Transport Coefficients}
\label{sec::reltrans}

In the absence of bound electron and partially-ionized impurity
physics, Ref.~\cite{Papp11a} generalizes the collision operator
coefficients $C_A$, $C_F$, $C_B$ to combine the non-relativistic
\cite{Trubnikov65} and relativistic \cite{Karney85} energy limits,
yielding
\begin{subequations}
\begin{align}
  C_A(v)&=\frac{\Gamma_{ee}\mathcal{G}(\frac{v} {v_{\rm
        th}})}{v}, \label{eq-CA}
  \\ C_F(v)&=\frac{\Gamma_{ee}\mathcal{G}(\frac{v}{v_{\rm
        th}})}{T_e}, \label{eq-CF} \\ C_B(v)&=
  \frac{\Gamma_{ei}}{2v}Z_{\rm eff} +\frac{\Gamma_{ee}}{2v}
  \Bigg[\erf\left(\frac{v}{v_{\rm
        th}}\right) \nonumber \\&-\mathcal{G}\left(\frac{v}{v_{\rm
        th}}\right)+\frac{1}{2}\left(\frac{v_{\rm
        th}v}{c^2}\right)^2\Bigg], \label{eq-CB}
\end{align}
\end{subequations}
where $\Gamma_{ee,ei}=n_ee^4\ln\Lambda_{ee,ei}/4\pi\epsilon^2_0$ with
$\ln\Lambda_{ee,ei}$ the Coulomb logarithm for e-e(e-i) collisions,
$v_{\rm th}=\sqrt{2T_e/m_e}$ the thermal electron velocity,
\begin{equation}
  \mathcal{G}(x)=\frac{\erf(x)-x\erf^{\prime}(x)}{2x^2}
\end{equation}
is the Chandrasekhar function, where the prime indicates a derivative
with respect to the independent variable. The expressions for
$\ln\Lambda_{ee,ei}$ are taken from Ref.~\cite{Hesslow17}
\begin{subequations}
\begin{align}
  \ln\Lambda_{ee}=&\ln\Lambda_0+\frac{1}{5}\ln\Big\{1+
  \left[2(\gamma-1)c^2/v_{\rm th}^2 \right]^{5/2} \Big\}
  \\ \ln\Lambda_{ei}=&\ln\Lambda_0+\frac{1}{5}\ln\left[1+
    \left(2\gamma v/v_{\rm th} \right)^5 \right]
  \\ \ln\Lambda_0=&14.9-\frac{\ln n_e(10^{20}\,{\rm m^{-3}})}{2}+\ln
  T({\rm keV}).
\end{align}
\end{subequations}
Note that the term in Eq.~(\ref{eq-CB}) proportional to $Z_{\rm eff}$
is taken from relativistic collision theory, and this term diverges as
$v\rightarrow 0$. For a $10\,{\rm MeV}$ RE, changing $T_e$ of the
companion plasma by a factor of $10$ yields a $6\%$ difference in
$\ln\Lambda_{ei}$, a $7\%$ difference in $\ln\Lambda_{ee}$, and factor
of 10 change to $\mathcal{G}$. However, in $C_F$, the change in
$\mathcal{G}$ is offset by the change in $T_e$, and in $C_B$,
$\mathcal{G}$ is small compared to the first term in the brackets,
leading to a low sensitivity of collisions on $T_e$ for REs. Conversely,
$C_B$ scales nearly as $(Z_{\rm eff}+1)$, so changes in $Z_{\rm eff}$
are strongly felt in the pitch angle scattering in the absence of
bound electrons.

Using the relation $dp = m_e\gamma^3dv$, the second
term in the brackets of Eq.~(\ref{eq-dp}) can be evaluated as
\begin{align}
  \frac{1}{p^2}\frac{\partial}{\partial p} \left[p^2C_A(p)\right]=&
  \frac{\Gamma_{ee}}{\gamma^3m_ev^2}\Bigg\{\left[2\gamma^2\left(
    \frac{v}{c}\right)^2-1\right]\mathcal{G}\left(\frac{v}{v_{\rm th}}\right)
  \nonumber \\ &+\frac{v}{v_{\rm
      th}}\erf^{\prime}\left(\frac{v}{v_{\rm th}}\right) \Bigg\}.
\end{align}
Note, that this form does not treat the momentum dependence in the
Coulomb logarithm.

\subsubsection{Bound Electrons}
\label{sec::bounde}

The transport coefficients given in Eqs.~(\ref{eq-CA}-\ref{eq-CB})
describe normalized collision frequencies due to Coulomb collisions
with ions and free electrons according to
\begin{subequations}
\begin{align}
  C_A(v)&\equiv
  \frac{p^2}{2}\nu_{\parallel}^{ee}, \label{eq-nupar}
  \\ C_F(v)&\equiv p\nu_{S}^{ee}, \label{eq-nus}
  \\ C_B(v)&\equiv \frac{p^2}{2}\left(\nu_{D}^{ei}
  +\nu_{D}^{ee}\right), \label{eq-nuD}
\end{align}
\end{subequations}
where the factor of $1/2$ used in Ref.~\cite{Trubnikov65} for the
definition of Eq.~(\ref{eq-nus}) has been absorbed into $\nu_S^{ee}$ as
done in Ref.~\cite{Hesslow17}. However the bound electrons of the
partially ionized impurities will also play an important
role. Reference \cite{Hesslow17} considers the effects of bound
electrons on the slowing down e-e collision frequency by including a
multiplicative factor according to Bethe's theory for collisional
stopping power \cite{Bethe30}
\begin{equation}
  \nu_S^{ee}=\nu_{S,CS}^{ee}\left\{1+\sum_j\frac{n_j}{n_e}
    \frac{Z_j-Z_{0j}}{\ln\Lambda}\left[\frac{1}{5}\ln(1+h_j^5)
    -\beta^2\right]\right\}, \label{eq-nuSH}
\end{equation}
where $\nu_{S,CS}^{ee}$ is the ``completely screened'' slowing down
frequency consistent with the models neglecting bound electron physics
given by Eqs.~(\ref{eq-nus},\ref{eq-CF}), the sum is over the
ionization state, $n_j$ is the density of the $j$-th ionization state,
$Z_j$ is the unscreened ({\it i.e.} fully ionized) impurity ion
charge, $Z_{0j}$ is the screened ( {\it i.e.}  partially ionized)
impurity ion charge, $h_j=p\sqrt{\gamma-1}/(m_ecI_j)$, where $I_j$ is the
mean excitation energy provided in Ref.~\cite{Sauer15}, and
$\beta=v/c$ is the usual relativistic speed. The factor $Z_j-Z_{0j}$
is recognized as the number of bound electrons a partially-ionized
impurity charge state possesses.  The following simulations assume
that all impurity charge states considered have the same
spatiotemporal profile as $n_e$, with only changes their ratio
$n_j/n_e$ satisfying
\begin{equation}
  n_e=\sum_jk_j n_j,\label{eq-chargestatemix}
\end{equation}
where $k_j$ is the charge state for a particular partially-ionized
impurity $j$.

The effects of bound electrons on the pitch-angle
diffusion e-i collision frequency is considered using the Born
approximation \cite{Kirillov75,Zhogolev14} to yield the modification
\begin{equation}
  \nu_{\rm D}^{ei}=\nu_{\rm D,CS}^{ei}\left(1+\frac{1}{Z_{\rm
      eff}}\sum_j \frac{n_j}{n_e}\frac{g_j}
     {\ln\Lambda_{ei}}\right), \label{eq-nuDH}
\end{equation}
where $\nu_{D,CS}^{ei}$ is the completely screened pitch angle scattering
frequency consistent with the models neglecting bound electron physics
given by Eq.~(\ref{eq-nuD}) and the first term of Eq.~(\ref{eq-CB}),
\begin{equation}
  g_j=\frac{2}{3}(Z_j^2\!-\!Z_{0j}^2)\ln(y_j^{3/2}\!+1\!)\!-\!\frac{2}{3}
  \frac{(Z_j\!-\!Z_{0j})^2y_j^{3/2}}{y_j^{3/2}\!+\!1}
\end{equation}
with $y_j=p\bar{a}_j/(m_ec)$, and $\bar{a}_j$ is the normalized effective ion
scale length for impurity charge state $j$. The values of $\bar{a}_j$ are
determined from the density of bound electrons as calculated in
Ref.~\cite{Hesslow18}. For a $10\,{\rm MeV}$ RE and a companion plasma
with $n_{Ne^+1}/n_{Ne^+2}=2$, the ratio between the $Z_{\rm eff}$ and
bound electron contributions to $\nu_D^{ei}$ is roughly $1.5:22.5$. If
the addition of Ar impurities from the initially injected pellet were
included, this ratio would most likely narrow, but still have the
bound electron contribution outweigh the $Z_{\rm eff}$
contribution to pitch angle scattering. Future modeling should better
account for more realistic partially-ionized impurity species.

With the model in Ref.~\cite{Hesslow18} of bound electrons and
partially-ionized impurities, the transport coefficients become
\begin{subequations}
\begin{align}
  C_{A,{\rm Bound}}(v)&=\frac{\Gamma_{ee}\mathcal{G}(\frac{v} {v_{\rm
        th}})}{v}, \label{eq-CAH}
    \\ C_{F,{\rm Bound}}(v)&=\frac{\Gamma_{ee}\mathcal{G}(\frac{v}{v_{\rm
        th}})}{T_e}\Bigg\{1+\sum_j\frac{n_j}{n_e}
  \frac{Z_j-Z_{0j}}{\ln\Lambda_{ee}} \nonumber
  \\ &\times\left[\frac{1}{5}\ln(1+h_j^5)
    -\beta^2\right]\Bigg\}, \label{eq-CFH} \\ C_{B,{\rm Bound}}(v)&=
    \frac{\Gamma_{ei}}{2v}Z_{\rm eff}\left(1+\frac{1}{Z_{\rm
        eff}}\sum_j \frac{n_j}{n_e}\frac{g_j} {\ln\Lambda_{ei}}\right)
    \nonumber \\ &+\frac{\Gamma_{ee}}{2v}
    \Bigg[\erf\left(\frac{v}{v_{\rm
          th}}\right)-\mathcal{G}\left(\frac{v}{v_{\rm
          th}}\right)+\frac{1}{2}\left(\frac{v_{\rm
          th}v}{c^2}\right)^2\Bigg]. \label{eq-CBH}
\end{align}
\end{subequations}

\begin{figure}
\centering
\noindent\includegraphics[width=3.4in]{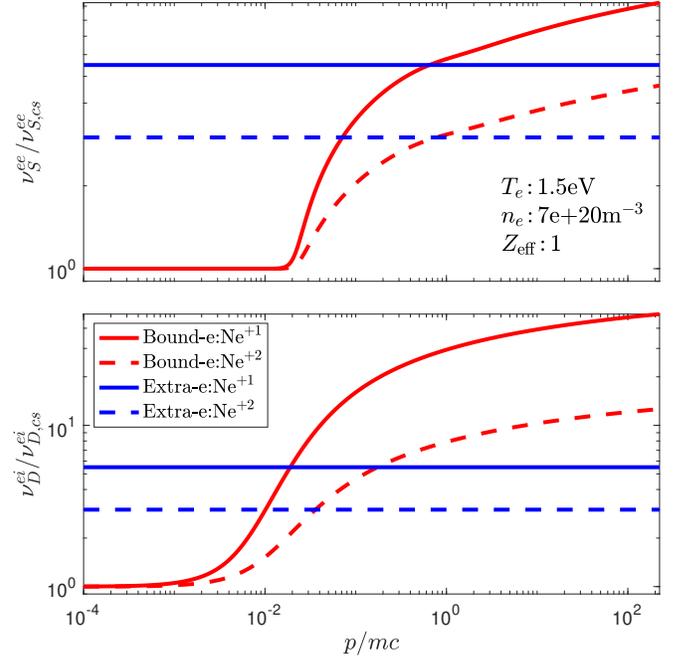}
\caption[]{\label{fig-bound}Top plot shows the slowing down collision
  frequency with the inclusion of bound electrons normalized to the
  frequency without bound electron physics. Bottom plot shows the
  pitch angle diffusion collision frequency with the inclusion of bound
  electrons normalized to the frequency without bound electron
  physics. Red traces indicate the Bound-e model and blue traces
  indicate the Extra-e model. Solid traces are for
  ${\rm Ne}^{+1}$ and dashed traces for for ${\rm Ne}^{+2}$.}
\end{figure}

Reference \cite{Hesslow17} also pointed to comments in
Ref.~\cite{Rosenbluth97} about the inclusion of bound electrons in
avalanche phenomena. Reference ~\cite{Hesslow17} incorporated these comments
by modifying the slowing down e-e collision frequency to include half
of the bound electrons
\begin{equation}
    \nu_S^{ee}=\nu_{S,CS}^{ee}\left(1+\sum_j\frac{n_j}{n_e}
      \frac{Z_j-Z_{0j}}{2}\right). \label{eq-nuSR}
\end{equation}
Note that bound electrons in Eq.~\ref{eq-nuSR} are treated the same as
free electrons when it comes to their effect on the collision
frequency, as compared with Eq.~\ref{eq-nuSH} that treats bound
electrons according to Bethe's theory for collisional stopping power.
We have also incorporated similar modifications to the pitch-angle
diffusion e-i collision frequency
\begin{equation}
    \nu_D^{ei}=\nu_{D,CS}^{ei}\left(1+\sum_j\frac{n_j}{n_e}
    \frac{Z_j-Z_{0j}}{2}\right). \label{eq-nuDR}
\end{equation}
Note that this modification only makes sense for the assumption
$Z_{\rm eff}=1$ being used in the present study. With the model
of bound electrons and partially-ionized impurities according to
Ref.~\cite{Rosenbluth97}, the transport coefficients become
\begin{subequations}
\begin{align}
  C_{A,{\rm Extra}}(v)&=\frac{\Gamma_{ee}\mathcal{G}(\frac{v} {v_{\rm
        th}})}{v}, \label{eq-CARP}
  \\ C_{F,{\rm Extra}}(v)&=\frac{\Gamma_{ee}\mathcal{G}(\frac{v}{v_{\rm
        th}})}{T_e}\left(1+\sum_j\frac{n_j}{n_e}
      \frac{Z_j-Z_{0j}}{2}\right), \label{eq-CFRP} \\ C_{B,{\rm Extra}}(v)&=
  \frac{\Gamma_{ei}}{2v}Z_{\rm eff}\left(1+\sum_j\frac{n_j}{n_e}
    \frac{Z_j-Z_{0j}}{2}\right) \nonumber
  \\ &+\frac{\Gamma_{ee}}{2v} \Bigg[\erf\left(\frac{v}{v_{\rm
        th}}\right)-\mathcal{G}\left(\frac{v}{v_{\rm
        th}}\right)+\frac{1}{2}\left(\frac{v_{\rm
        th}v}{c^2}\right)^2\Bigg]. \label{eq-CBRP}
\end{align}
\end{subequations}

In the following simulations, we will refer to 3 collision models,
``No-Bound'' with transport coefficients given by
Eqs.~(\ref{eq-CA}-\ref{eq-CB}) with no bound electrons, ``Bound-e''
with transport coefficients given by Eqs.~(\ref{eq-CAH}-\ref{eq-CBH}),
and ``Extra-e'' with transport coefficient given by
Eqs.~(\ref{eq-CARP}-\ref{eq-CBRP}). Figure \ref{fig-bound} shows the
slowing down frequency in the top plot and the pitch angle scattering
frequency in the bottom plot for singly-ionized Ar and Ne using the
Bound-e and Extra-e models. The collision frequencies are normalized to
the completely screened, or No-Bound, collision frequencies.

\subsection{Radiation Damping}

Simulations also include the effects of synchrotron radiation due to
the radiation reaction force ${\bf F}_R$.  The Landau-Lifshitz
representation \cite{Landau71} of the Lorentz-Abraham-Dirac radiation
reaction force, ignoring the electric field and advective derivatives,
is
\begin{equation}
  {\bf F}_R=\frac{1}{\gamma\tau_R}\left[\left({\bf p}\times{\bf
      b}\right)\times{\bf b}-\frac{1}{(m_ec)^2}\left({\bf p}\times{\bf
      b}\right)^2{\bf p}\right], \label{eq-LLLAD}
\end{equation}
where $\tau_R=6\pi\epsilon_0(m_ec)^3/(e^4B^2)$ is the radiation
damping time scale.  In azimuthally-symmetric spherical momentum
space, with the identity $(\hat{p}\times{\bf b})\times{\bf
  b}=-\sin\eta\hat{p}-\cos\eta\hat{\eta}$, the flux-conserving
form of the Fokker-Planck PDE can be written as two equivalent SDEs
given by
\begin{subequations}
  \begin{align}
    \frac{dp}{dt}&=-\frac{\gamma p}{\tau_R}(1-\xi^2) \label{eq-dpdt}\\
    \frac{d\xi}{dt}&=\frac{\xi(1-\xi^2)}{\tau_R\gamma}. \label{eq-dxidt}
  \end{align}
\end{subequations}
Equation (\ref{eq-dpdt}) is consistent with the relativistic Larmor
formula ({\it e.g.} \cite{Jackson62}).

The evolution equations for $(p,\xi)$ in
Eqs.~(\ref{eq-dpdt}-\ref{eq-dxidt}) can be transformed into evolution
equations for $(p_\parallel,\mu)$ yielding
\begin{subequations}
  \begin{align}
  \frac{dp_\parallel}{dt}&=-\frac{p_\parallel(1-\xi^2)}{\tau_R}
  \left(\gamma-\frac{1}{\gamma}\right) \label{eq-SRp}
  \\ \frac{d\mu}{dt}&=-\frac{2\mu}{\tau_R}\left[\gamma(1-\xi^2)+
    \frac{\xi^2}{\gamma}\right]. \label{eq-SRxi}
\end{align}
\end{subequations}
These deterministic evolution equations are added to the Cash-Karp
algorithm for integrating the GC equations of motion by adding the RHS
of Eq.~(\ref{eq-SRp}) to the RHS of Eq.~(\ref{eq-gcp}) and using
Eq.~(\ref{eq-SRxi}) to evolve the magnetic moment. We note that
Eq.~(\ref{eq-LLLAD}) is to be taken at the location of a moving
charge, so the above implementation is only valid for a small
gyroradius. A specific formulation for the GC equations of motion is
given by Ref.~\cite{Hirvijoki15} and will be explored in future
studies.

The bremsstrahlung radiation due to runaway electrons interacting with
impurities modeled as a stopping power is discussed in
Ref.~\cite{Bakhtiari04} and can be written as
\begin{align}
  \frac{d}{dt}\left[(\gamma-1)m_ec^2\right]&=-2vn_j\kappa
  Z_{0j}(Z_{0j}+1) \nonumber
  \\ &\times\frac{\alpha}{\pi}(\gamma-1)\left[\ln(2\gamma)-\frac{1}{3}\right],
\end{align}
where $\kappa=2\pi r_e^2m_ec^2$, $r_e=e^2/4\pi\epsilon_0m_ec^2$ is the
classical electron radius, and $\alpha=1/137$ is the fine structure
constant. Because only the energy changes, and not the pitch angle,
using $d\gamma/dp=v/(m_ec^2)$ we can write
\begin{subequations}
\begin{align}
  \frac{dp}{dt}&=-2n_j\kappa Z_{0j}(Z_{0j}+1)
  \nonumber
  \\ &\times\frac{\alpha}{\pi}(\gamma-1)\left[\ln(2\gamma)-\frac{1}{3}\right]\\ \frac{d\xi}{dt}&=0.
\end{align}
\end{subequations}
Similarly as for synchrotron radiation, the evolution equations for
$(p,\xi)$ can be transformed into evolution equations for
$(p_\parallel,\mu)$ yielding
\begin{subequations}
\begin{align}
  \frac{dp_\parallel}{dt}&=-\xi2n_j\kappa Z_{0j}(Z_{0j}+1) \nonumber
  \\ &\times\frac{\alpha}{\pi}(\gamma-1)\left[\ln(2\gamma)-
    \frac{1}{3}\right] \label{eq-BRp}\\ \frac{d\mu}{dt}&=
  -\frac{(1-\xi^2)p}{m_eB}2n_j\kappa Z_{0j}(Z_{0j}+1)\nonumber
  \\ &\times\frac{\alpha}{\pi}(\gamma-1)\left[\ln(2\gamma)
    -\frac{1}{3}\right] \label{eq-BRxi}.
\end{align}
\end{subequations}
These deterministic evolution equations are also added to the
Cash-Karp algorithm for integrating the GC equations of motion by
adding the RHS of Eq.~(\ref{eq-BRp}) to the RHS of Eq.~(\ref{eq-gcp})
and adding the RHS of Eq.~(\ref{eq-BRxi}) to the RHS of
Eq.~(\ref{eq-SRxi}). We note that Ref.~\cite{Embreus16} formulates a
Boltzmann collision operator for bremsstrahlung radiation that results
in a more accurate momentum-space evolution compared to stopping power
models. However, for the present study concerning RE current
dissipation, the precise details of momentum phase space evolution
are of lesser importance.

\section{Ensemble Initialization}
\label{sec::init}

A Metropolis-Hastings (MH) algorithm \cite{Metropolis53,Hastings70} is
employed to sample user-provided, multidimensional distribution
functions using Markov-chains of Gaussian processes in each
dimension. The ``acceptance ratio'' is calculated for every sample of
the Markov-chains, comparing the new sample to the previous sample. If
this ratio is greater than $1$ or a uniformly-distributed random
number, the new sample is accepted. Included in the acceptance ratio
is the Jacobian determinant of the cylindrical spatial coordinate
system $R$ and the spherical momentum phase space $p^2\sin\eta$.  One benefit of the MH algorithm is that it only
depends on the ratio of the distribution function at different
sampling points, and not the absolute values, thus no normalizations
are needed.

\begin{figure}
\centering
\noindent\includegraphics[width=3.4in]{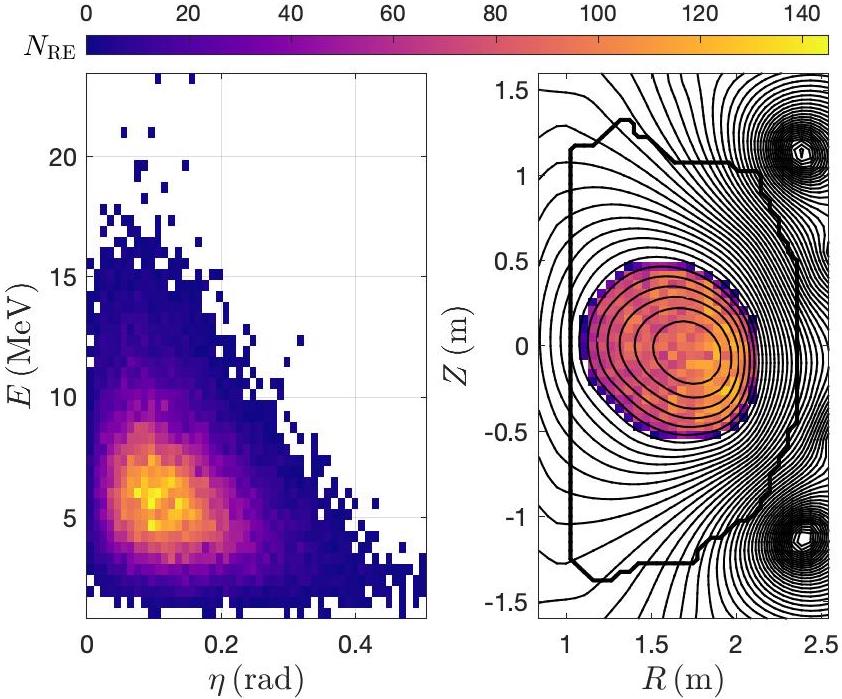}
\caption[]{\label{fig-16init}Example initial distribution used for
  simulations of DIII-D discharge \#164409 with Ne MGI with
  $2.5\times10^4$ sampled particles. $E,\eta$ distribution shown in
  a), and $R,Z$ distribution shown in b), with overlaid (thin)
  contours of $\psi_p$ and (thick) first wall.}
\end{figure}

\begin{figure}
\centering
\noindent\includegraphics[width=3.4in]{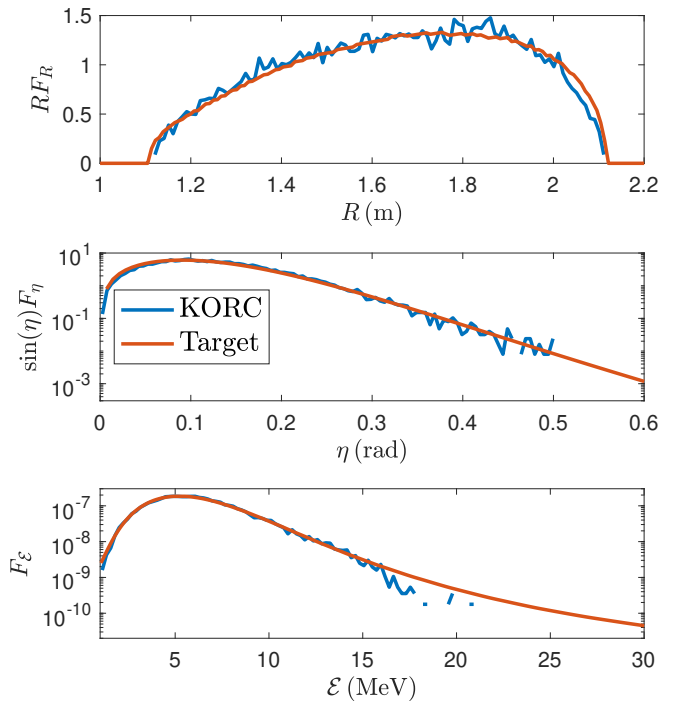}
\caption[]{\label{fig-16sample}Comparison between the target
  distribution (red traces) compared to the sampled distribution (blue
  traces) for a given KORC calculation. The top plot shows the spatial
  distribution for a uniform density of REs for DIII-D discharge
  \#164409. The middle plot shows the pitch angle distribution given
  by Eqs.~\ref{eq-pitch1},\ref{eq-pitch2}. The bottom plot shows the
  energy distribution for sampling of the distribution function
  inferred from a separate DIII-D discharge reported in
  Ref.~\cite{Hollmann15a}.}
\end{figure}

For the following simulations, we use an initial distribution of the
form
\begin{equation}
  f[R,Z,\mathcal{E},\eta]=f_{\psi_p}[\psi_p(R,Z)]f_{\mathcal{E}}(\mathcal{E})
  f_\eta(R,\mathcal{E},\eta),
\end{equation}
where $\mathcal{E}=\gamma m_ec^2$ is the total energy of a RE. Across
all simulations to be presented, we assume that the initial spatial
distribution has a Gaussian dependence of the form
\begin{equation}
  f_{\psi_p}(R,Z)=\exp\left[-\frac{\psi_N(R,Z)}{\sigma_{\psi_N}}\right],
\end{equation}
where $\psi_N=[\psi_p(R,Z)-\psi_{p,{\rm axis}}]/(\psi_{p,{\rm
    lim}}-\psi_{p,{\rm axis}})$ is the normalized poloidal flux, with
$\psi_{p,{\rm axis}}=0.600\,{\rm Wb}$ the initial poloidal flux at the
magnetic axis and $\psi_{p,{\rm lim}}=0.845\,{\rm Wb}$ is the initial
poloidal flux at the HFS limiter. Note that this parameterization is
chosen because we desire a spatial distribution as a function of minor
radius $r$, and $\psi~r^2$.  A uniform distribution can be recovered
by choosing $\sigma_{\psi_N}$ arbitrarily large, {\it e.g.}
$\sigma_{\psi_N}=10^6$ , and an indicator function is used to limit
sampling to where $\psi_N<0.845$. The particles are sampled uniformly
in toroidal angle $\phi$. For $f_{\mathcal{E}}(\mathcal{E})$ we use
either a monoenergetic distribution of $10\,{\rm MeV}$ or the
distribution inferred from experimental data of a post-disruption, RE
beam in Ref.~\cite{Hollmann15a}. For the pitch angle distribution we
use either a monopitch distribution of $10^\circ$ or that considered
in Ref.~\cite{Hollmann15a}, which assumes a balance between pitch
angle scattering and electric field ``pinching''
\begin{equation}
  f_\eta(R,\mathcal{E},\eta)=\frac{A(R,\mathcal{E})}{2\sinh
    A(R,\mathcal{E})}e^{A(R,\mathcal{E})\xi},\label{eq-pitch1}
\end{equation}
with
\begin{equation}
  A(R,\mathcal{E})=\frac{2\tilde{E}_\phi(R)}
  {Z_{\rm eff}+1}\frac{\gamma^2-1}{\gamma},\label{eq-pitch2}
\end{equation}
where $\tilde{E}_\phi(R)=E_0R_0/(E_{\rm CH}R)$ is the normalized,
initial, toroidal electric field, with the Connor-Hastie field
\cite{Connor75} $E_{\rm
  CH}=n_ee^3\ln\Lambda_0/4\pi\epsilon_0^2m_ec^2$, magnitude
$E_0/E_{\rm CH}=24.56$, and $Z_{\rm eff}=1$ is effective impurity
nuclear charge.

For the energy and pitch angle distribution from Fig. 5
Ref.~\cite{Hollmann15a}, the resulting distribution function is shown
in Fig.~\ref{fig-16init} as a function of a) $E,\eta$ and b) $R,Z$ for
$2.5\times10^4$ particles. Note that Fig.~\ref{fig-16init}b) has
overlaid (thin line) contours of $\psi_p$, consistent with the initial
time in Sec.~\ref{sec-plasmamodel}, and indicates the (thick line)
approximate first wall on DIII-D. The apparent increase in the number
of REs as $R$ increases reflects the Jacobian determinant of the
cylindrical coordinate system appearing when we integrate over the
$\phi$ direction to show the distribution in the $R,Z$ plane.

A comparison between the target distribution and sampled distribution
is presented in Fig.~\ref{fig-16sample}. The top plot shows the
distribution $F_R=\int f_{\psi_p}dZ$, where the factor of $R$ is the
Jacobian determinant of the cylindrical coordinate system used for the
spatial representation. The middle plot shows the distribution
$F_\eta=\int dZ\int RdR\int d\mathcal{E} f$, where the factor of
$\sin(\eta)$ comes from the Jacobian determinant of the spherical
coordinate system used for the momentum representation. And lastly the
bottom plot shows the distribution $F_{\mathcal{E}}=f_{\mathcal{E}}$
reported in Ref.~\cite{Hollmann15a}. Note that because the target
$F_{\mathcal{E}}$ decreases algebraically, rather than exponentially,
the Gaussian process MH algorithm has difficulties sampling the high
energy tail. However, since the majority of the RE beam energy is
contained in the bulk around $6.5\,{\rm MeV}$, this sampling is
acceptable for the present modeling.

\section{Results}
\label{sec::results}

\begin{table}[]
  \centering
  \begin{tabular} {c|c|c|c|c|c}
    Simulation & Collisions & Bound model & $E_\phi$ & ST model &
    $f(\mathcal{E},\eta)$ \\ \hline Case 1 & F & NA & T &
    NA & mono-$\mathcal{E},\eta$ \\ \hline Case 2 & F & NA & F  &
    NA & mono-$\mathcal{E},\eta$ \\ \hline Case 3 & T & Bound-e & T &
    diffuse & mono-$\mathcal{E},\eta$ \\ \hline  Case 4 & T & Extra-e & T &
    diffuse & mono-$\mathcal{E},\eta$ \\ \hline  Case 5 & T & No-bound & T &
    diffuse & mono-$\mathcal{E},\eta$ \\ \hline  Case 6 & T & Bound-e & T &
    diffuse & Ref. \cite{Hollmann15a} \\ \hline  Case 7 & T & Bound-e & T &
    dense & mono-$\mathcal{E},\eta$ \\ \hline  Case 8 & T & Bound-e & T &
    dense & Ref. \cite{Hollmann15a} \\       
  \end{tabular}
  \caption[]{\label{tab-sims}Models used in simulations shown in this
    section. 'F/T' are abbreviations for false/true.}
\end{table}

To characterize the evolution of the ensemble of particles at a
macroscopic level, we define the total RE energy $\mathcal{E}_{\rm RE}$
\begin{equation}
  \mathcal{E}_{\rm RE}(t)=m_ec^2\sum_i^{N_p}
  \gamma_i(t)\mathcal{H}_{{\rm RE},i}(t) 
\end{equation}
where
\[
  \mathcal{H}_{{\rm RE},i}(t) =
  \begin{cases}
                                   1 & \text{if $p_i(t)>m_ec$} \\ 0 &
                                   \text{if $p_i(t)<m_ec$ or hits
                                     wall.}
  \end{cases}
\]
$\mathcal{H}_{\rm RE}$ effectively splits the ensemble of particles
into two populations, confined and thermalized or deconfined
electrons. Note that this definition is consistent with the discussion
at the end of Sec.~\ref{sec-collisions}, and roughly consistent with
the definition of critical momentum in Ref.~\cite{Rosenbluth97}. We
also define the RE current $I_{\rm RE}$, beginning from the toroidal
current density $J_\phi=-en_e{\bf v}\cdot\hat{\phi}$, satisfying
$I_{\rm RE}=\int dR\int dZ J_\phi$. Using the definition of the
density of an ensemble of particles $n_e({\bf x})=\sum_i\delta^3({\bf
  x}-{\bf x}_i)=\sum_i\delta(R-R_i)\delta(Z-Z_i)/(2\pi R)$ and
assuming that the toroidal component of the velocity is, to lowest
order, ${\bf v}\cdot\hat{\phi}\simeq v_\parallel{\bf
  b}\cdot\hat{\phi}=v\xi b_\phi$ yields
\begin{equation}
  I_{\rm RE}(t)=-\frac{e}{2\pi}\sum_i^{N_p} \frac{v_i\xi_i
      b_{\phi,i}}{R_i} \mathcal{H}_{{\rm RE},i}(t). \label{eq-Ire}
\end{equation}
To make comparisons between simulations and experiments we normalize
these quantities to their initial values $\mathcal{E}_{\rm RE}(0)$ and
$I_{\rm RE}(0)$. The initial values correspond to time $1.405\,{\rm
  ms}$ from DIII-D discharge \#164409.

The following calculations all use dynamic magnetic and electric
fields unless otherwise noted, and synchrotron and bremsstrahlung
radiation. We note that the role of radiation for the present study of
RE dissipation by impurity injection is negligible (not shown). Table
\ref{tab-sims} summarizes all the simulations presented in this
section, with the combination of collisional, bound electron, electric
field, spatiotemporal electron and impurity density, and initial RE
energy and pitch distribution models used.

\subsection{Confinement Losses}
\label{sec::confinement}

\begin{figure}
\centering
\noindent\includegraphics[width=3.4in]{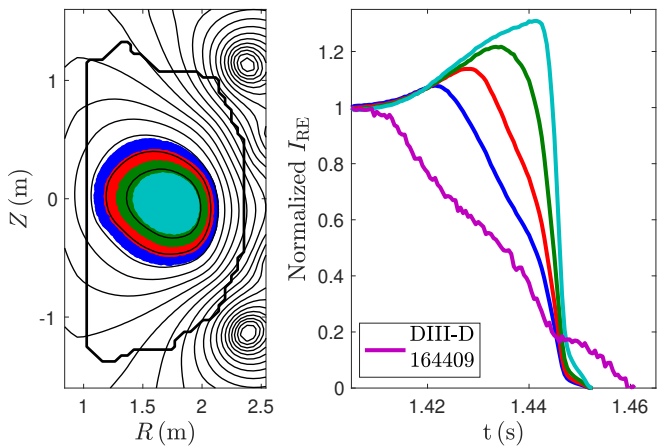}
\caption[]{\label{fig-beamwidthloss}Subsets of REs, delineated by
  initial value of poloidal flux function consistent with the left
  figure, to study the effect of different beam widths, made possible
  by the linear nature of individual RE orbits. Calculation performed
  consistent with Case 1 model.}
\end{figure}

We begin by simulating a RE beam without any collisions, which
decouples the effects of deconfinement of REs due to the evolution of
the experimentally-reconstructed fields from collisional
effects. Fig.~\ref{fig-beamwidthloss} shows such a simulation, where
we indicate different initial RE beam widths in the left plot,
including particles within $\psi_N<1.0$ in dark blue, $\psi_N<0.78$ in
red, $\psi_N<0.56$ in green, and $\psi_N<0.34$ in light blue. This is
accomplished by a single simulation with a spatially uniform RE
distribution, because of the linear nature of RE orbits. The right
plot in Fig.~\ref{fig-beamwidthloss} shows that as the initial RE
beam width gets smaller, the effect of deconfinement is delayed.
Physically, the smaller radii beams will not interact with the inner
wall until later times, but then will have all their particles
deconfined rapidly. The most extreme case would be a pencil beam,
which would not lose any particles until the beam touches the wall,
and then all particles would be lost nearly instantaneously. The
normalized $I_{\rm RE}$ for each subset of simulated REs is compared
to the experimental current from DIII-D discharge \#164409, indicated
by the violet trace.

The increase in normalized RE current is an interesting result. As the
magnetic configuration evolves, the magnetic axis advects toward the
HFS, decreasing the major radial location of REs summed in
Eq.~(\ref{eq-Ire}). Also noted in Ref.~\cite{Eidietis12}.  This makes
intuitive sense, as all REs are approximately traveling at $c$ in the
toroidal direction, starting at a pitch of $\eta=10^\circ$ with a
small spread in pitch angle due to spatial orbit effects
\cite{Carbajal17a}. Without collisions, this speed remains the same,
and as the RE beam advects toward the HFS, where the toroidal orbit
length decreases linearly with $R$, the toroidal transit distance, and
therefore time, becomes less, yielding more $I_{\rm RE}$.  There will
also be an additional effect due to pitch angle ``pinching'' due to
the toroidal electric field, but we find that this is a small effect
compared to the major radial location (not shown).

At the time after all REs are either deconfined or thermalized in this
and each of the following simulations, DIII-D discharge \#164409 shows
current remaining, approximately $46\,{\rm kA}$, for an additional
$10\,{\rm ms}$.  With the present modeling capabilities of KORC, we
are unable to study this discrepancy directly, and will address it in
future work.  However, possible reasons for the remaining current
include a possible lack of magnetic reconstruction fidelity, secondary
REs generated by large-angle collisions during this time period of
larger induced toroidal electric field, or evolution of the companion
plasma. First, from the poloidal flux contours in the final plot of
Fig.~\ref{fig-psiE} for $t=1.445\,{\rm s}$, there is an open magnetic
configuration, consistent with the loss of nearly all REs at this
time, except for those trapped on the low field side.  It is possible
that JFIT, which is only constrained to external magnetic diagnostics,
could be incorrectly reconstructing the internal magnetic
configuration, and a closed flux region could be remaining.  Second,
if our modeling contained a secondary RE source, these added REs would
have a high probability to be generated in the trapped region, and
subsequently confined. Third, from Fig.~\ref{fig-denevocompare} there
is still a significant line integrated electron density being observed
on the V1 chord in the open flux region, indicating that plasma is not
being rapidly deconfined as one may expect. For a crude estimate, the
companion plasma could be carrying approximately $4.2\,{\rm kA}$ to
$43\,{\rm kA}$ of Ohmic current, for a range of electron temperature
of $1.5\,{\rm eV}$ to $8\,{\rm eV}$, possibly making a significant
contribution to the observed current. Additional study is required to
better understand this final time interval of the RE dissipation.

\subsection{Bound electron modeling}
\label{sec::bound}

\begin{figure}
\centering
\noindent\includegraphics[width=3.4in]{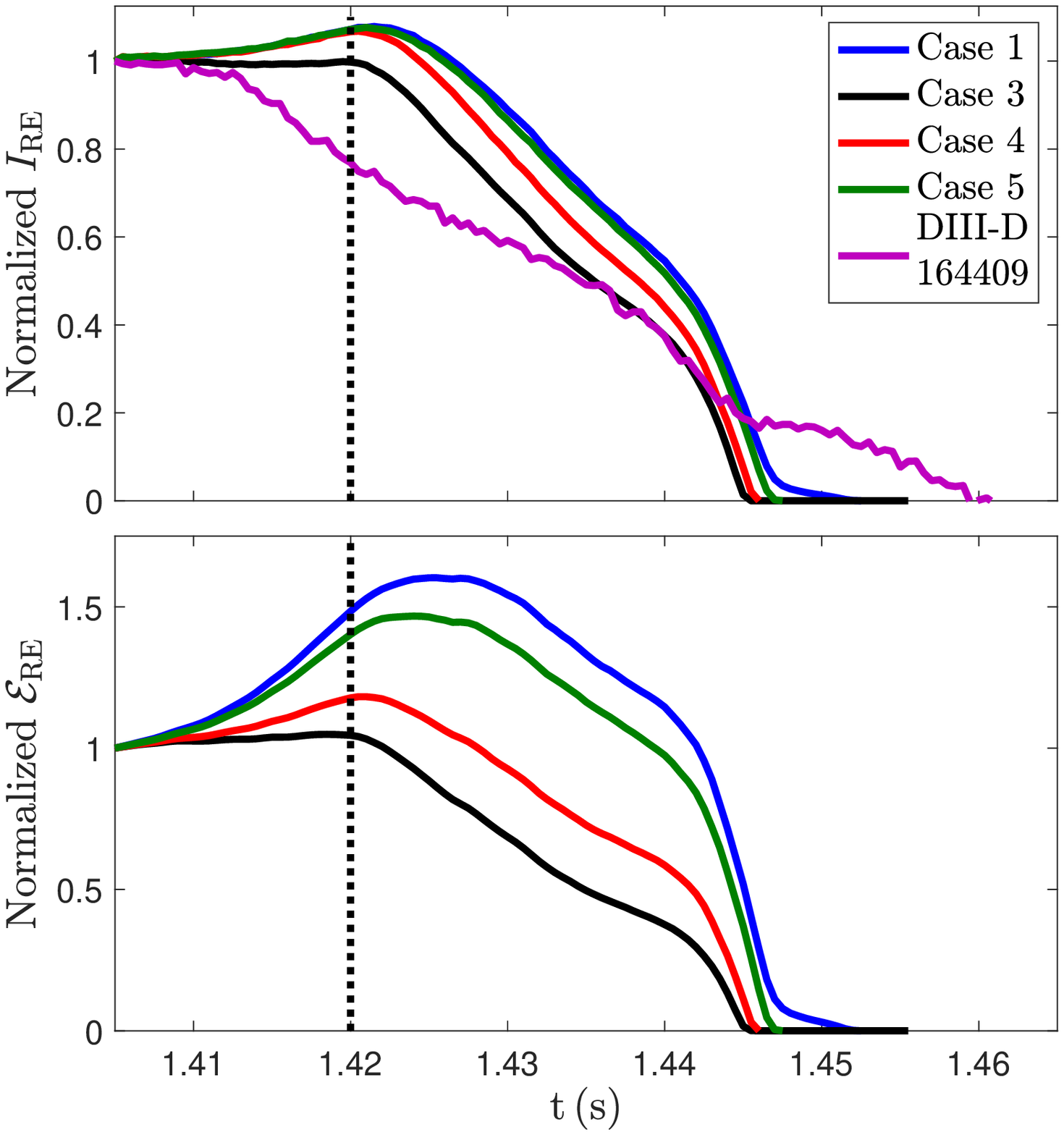}
\caption[]{\label{fig-boundmodels}Comparison of evolution of
  normalized RE beam current (top plot) and RE beam energy (bottom
  plot) for different bound electron models indication in Table
  \ref{tab-sims} with DIII-D discharge \#164409 indicated by the
  violet trace. The vertical, dashed, black line indicated the
  approximate time that deconfinement begins to play a role, as
  estimated from Fig.~\ref{fig-ConVsCol}.}
\end{figure}

\begin{figure}
\centering
\noindent\includegraphics[width=3.4in]{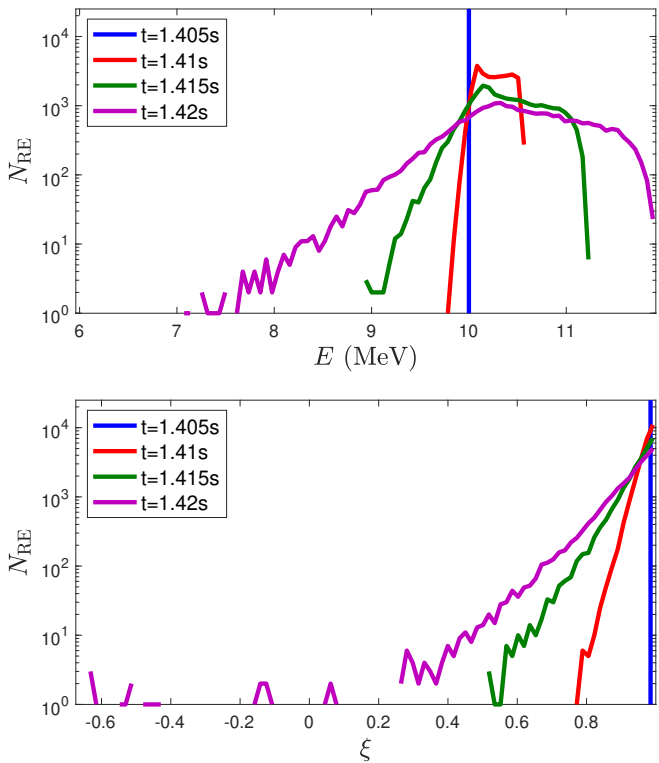}
\caption[]{\label{fig-histEeta_evo}Evolution of the RE ensemble energy
  (top plot) and cosine of the pitch angle (bottom plot) for the
  simulation in Fig.~\ref{fig-boundmodels} using the Case 3
  model. Traces show the distributions every $5\,{\rm ms}$ until
  confinement begins to degrade as estimated from
  Fig.~\ref{fig-ConVsCol}.}
\end{figure}

As a next step we include Coulomb collisions, and investigate the
effect of different models of bound electron physics developed in
Secs.~\ref{sec::reltrans} and \ref{sec::bounde}. Figure
\ref{fig-boundmodels} shows the dependence of the normalized RE
current (top plot) and energy (bottom plot) evolution on the bound
electron model.  The blue trace corresponds Case 1 shown as the dark
blue trace in Fig.~\ref{fig-beamwidthloss}, the black trace uses Case
3 with the Bound-e model from Eqs.~\ref{eq-CAH}-\ref{eq-CBH}, the red
trace uses Case 4 with the Extra-e model from
Eqs.~\ref{eq-CARP}-\ref{eq-CBRP}, the green trace uses the Case 5 with
the No-Bound electron model from Eqs.~\ref{eq-CA}-\ref{eq-CB}, and the
violet trace is the current from DIII-D discharge \#164409. The
vertical, dashed, black line indicated the approximate time that
deconfinement begins to play a role, as estimated from
Fig.~\ref{fig-ConVsCol}.  Progressing from no collisions to the
No-Bound, Extra-e, and lastly Bound-e collision models, the KORC
simulated normalized RE current more closely aligns with the
experimental current.  We note that while RE current evolution
measurements are routine in tokamak experiments, evolution of the RE
energy is not presently available in all discharges. Recent results in
Ref.~\cite{Lvovskiy20} show the evolution of the energy distribution,
as observed using the Gamma Ray Imager
\cite{Cooper16,Pace16}. However, these measurements presently require
a low level of HXR bremsstrahlung flux, and are not available in
discharges with a high level of high-Z impurities. Future work
comparing KORC simulations with discharges from this scenario would
make for a good validation study.

While nearly all the KORC simulations have their normalized RE current
evolve qualitatively similarly, there is great disparity in how the
normalized RE energy evolves in each case. Each model has the energy
increasing until after confinement is lost, with the main difference
being the rate at which energy increases due to acceleration by the
toroidal electric field in each case. The collisional slowing down is
nonexistent in the no collisions simulation and is so weak in the No
Bound model, that both cases still have energy increasing after loss
of confinement. For these parameters, it is striking how the No-Bound
electron model is barely different from no collision case, pointing to
the conclusion that the inclusion of a bound electron model is
essential for recovering accurate simulation results. The simulation
incorporating the Bound-e model is the only one to have any current
dissipate before loss of confinement, albeit at the small amount of
$>1\%$, and the least energy increase of $4.5\%$ The simulation using
the Extra-e model yields similar results to that using the
Bound-e model, but due to greatly increased pitch angle scattering, by
a factor of $7.15$ at $10\,{\rm MeV}$, and marginally increased
collisional friction, by a factor of $1.33$ at $10\,{\rm MeV}$, in
Bound-e model, as seen in Fig.~\ref{fig-bound}.

To better understand the RE ensemble averaged energy and current, it
is instructive to directly view the evolution of the energy and pitch
angle distributions. Figure \ref{fig-histEeta_evo} shows the evolution
of the energy (top plot) and cosine of the pitch angle (bottom plot)
for the simulation using the Bound-e model. Before REs begin to be
deconfined, it is clear that the average energy is increasing, while
$\xi$ is decreasing, or rather $\eta$ is increasing. We view $\xi$ as
compared to $\eta$, because $\xi$ is a direct input into the
calculation of the current consistent with Eq.~(\ref{eq-Ire}). We
posit that the energy is increasing due to REs being accelerated by
the induced toroidal electric field more than decelerated by
collisional slowing down. Comparing the approximate time rate of
change of the momentum due to the induced toroidal field
$dp/dt=-eE_\phi\xi$ and collisional slowing down for the initial RE
energy and pitch, the electric force is initially equal to the
collisional slowing down and increases as a larger toroidal electric
field is induced. Conversely, we hypothesize that the pitch angle is
increasing due to collisional pitch angle scattering with the Bound-e
model greater than the pinching effect due to the induced toroidal
electric field. Comparing the time rate of change of the cosine of the
pitch angle due to the induced toroidal field
$d\xi/dt=-eE_\phi(1-\xi^2)/p$ and collisional pitch angle scattering
for the initial RE energy and pitch, the collisional pitch angle
scattering is approximately $10^2$ larger than the electric force
pinching effect.

\subsection{Parametric Modeling}
\label{sec::parametric}

\begin{figure}
\centering
\noindent\includegraphics[width=3.4in]{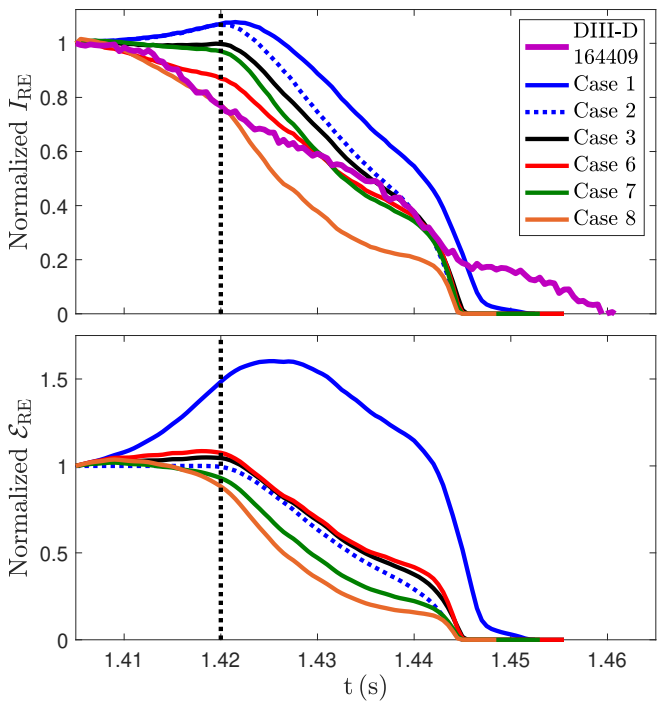}
\caption[]{\label{fig-2x1t}Comparison of evolution of normalized RE
  beam current (top plot) and RE beam energy (bottom plot) for a
  hierarchy of models indicated in Table \ref{tab-sims} with DIII-D
  discharge \#164409 indicated by the violet trace. The vertical,
  dashed, black line indicated the approximate time that deconfinement
  begins to play a role, as estimated from Fig.~\ref{fig-ConVsCol}.}
\end{figure}

Now that we have discussed the effects of deconfinement and bound
electron models, we turn our attention to the parameterization of the
initial RE distribution and spatiotemporal electron and
partially-ionized impurity profile. Figure \ref{fig-2x1t} shows the
dependence of the normalized RE current (top plot) and energy evolution
(bottom plot) on the initial RE distribution and spatiotemporal density
profile. The dark blue trace is Case 1 without collisions shown
as the dark blue trace in Fig.~\ref{fig-beamwidthloss}; the dotted,
dark blue trace is Case 2 without collisions or the dynamic toroidal
electric field; the black trace is the canonical Case 3 using the
Bound-e model shown in Fig.~\ref{fig-boundmodels}; the red trace is
Case 6 that uses the initial RE energy and pitch distribution from
Ref.~\cite{Hollmann15a}; the green trace is Case 7 using the narrow
and dense spatiotemporal density profile; and the orange trace is Case
8 using both the initial RE energy and pitch distribution from
Ref.~\cite{Hollmann15a} and the narrow and dense spatiotemporal
density profile.

The collisionless simulation without the induced toroidal electric
field (Case 2) indicates that the effect of the toroidal electric
field nearly balances out collisional dissipation, as it is nearly
identical to the canonical case. This reinforces our estimation of the
comparison of the two forces from the end of Sec.~\ref{sec::bound}.

It can be seen that varying the initial energy and pitch distribution
has a significant effect on current dissipation but not energy
dissipation. We postulate that this is due to lower initial mean energy,
which yields a larger pitch angle scattering consistent with
Eq.~(\ref{eq-deta}) or (\ref{eq-dxi}) that varies as $1/p^2\sim 1/E^2$.

Varying spatiotemporal density profile to use the narrow and dense profile
has a marginal effect on current dissipation but significant effect on
energy dissipation. We posit that this is due to the collisional force
being approximately twice as large as the toroidal electric force for
the initial conditions for the more dense spatiotemporal density
profile, whereas it was approximately equal for the more diffuse
spatiotemporal density profile. By itself, varying the spatiotemporal
density profile decreases the normalized energy before deconfinement,
but only by $7\%$.

When the effects of the different initial energy and pitch
distribution and denser spatiotemporal density are combined, both the
current and energy are dissipated to a higher degree than with either of
the effects separately. This can be viewed as the best case scenario
from the modeling of RE mitigation via Ne MGI in DIII-D discharge
\#164409. There is, however, a shortfall in the simulated current
after confinement degrades for the case with combined effects. We
hypothesize that this shortfall would be augmented by the evolution of the
companion plasma or secondary REs generated by large-angle collisions
during this time period of larger induced toroidal electric field. As
was the case in Sec.~\ref{sec::confinement}, we will discuss this
further in Sec.~\ref{sec::conclusion}.

\begin{figure}
\centering
\noindent\includegraphics[width=3.4in]{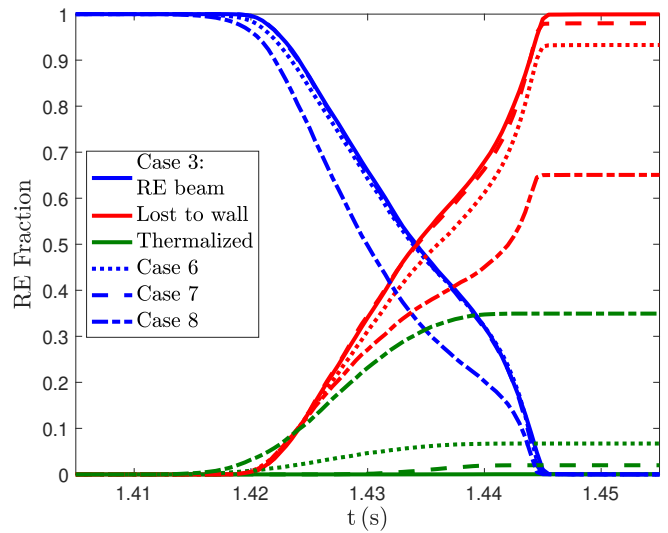}
\caption[]{\label{fig-ConVsCol}Comparison of the time evolution of the
  number of confined REs (blue traces), deconfined REs (red traces),
  and collisionally dissipated REs (green traces), where the compared
  simulations use models given in Table \ref{tab-sims}.}
\end{figure}

We can also view the evolution of particles in the RE beam, deconfined
particles impacting the wall, and thermalized particles whose momentum
fall below $p<m_ec$. Figure \ref{fig-ConVsCol} shows the evolution of
particles in the RE beam (blue traces), deconfined particles (red
traces), and thermalized particles (green traces) for simulations
presented in Fig.~\ref{fig-2x1t}, where the canonical simulation is
indicated with solid traces, the simulation with initial RE energy
and pitch distribution from Ref.~\cite{Hollmann15a} indicated by the
dotted traces, the simulation with the more dense spatiotemporal
density profile indicated by the dashed traces, and the simulation
combining both effects by the dash-dotted traces.

The previous results are borne out clearly in Fig.~\ref{fig-ConVsCol},
namely that the majority of REs across all simulations are lost to the
wall, rather than thermalized. The canonical simulation (Case 3) is the most
dire situation with $>0.1\%$ thermalized, the dense ST simulation has
$2\%$ thermalized, the simulation with initial RE energy and pitch
distribution from Ref.~\cite{Hollmann15a} has $7\%$ thermalized, and
the simulation with combined effects is again the best case scenario
with $35\%$ thermalized.  The simulations using initial RE energy
distribution from Ref.~\cite{Hollmann15a} have lower average initial
distribution energy, where the collisional slowing down is more
effective.

\subsection{Additional Experimental Connections}
\label{sec::expconnect}

\begin{figure}
\centering
\noindent\includegraphics[width=3.4in,bb=24 3 512 372,clip=true]{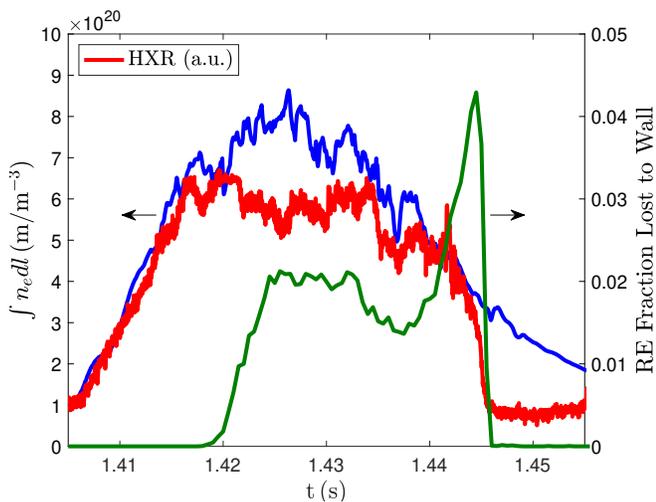}
\caption[]{\label{fig-HXRvsDen}Comparison between the V1
  interferometer diagnostic (blue trace) and HXR signal (red trace)
  for DIII-D discharge \#164409 with the RE fraction lost to the wall
  for the canonical KORC calculation (solid green trace).}
\end{figure}

In Fig.~\ref{fig-d3draw}d, the HXR emission is observed to increase
rapidly upon the injection of Ne gas into the post-disruption RE
beam. The HXR signal (red trace) has been reproduced in
Fig.~\ref{fig-HXRvsDen}, and plotted with the line integrated electron
density from vertical chord V1 (blue trace) and the simulated RE
fraction lost to the wall for the canonical KORC calculation (green
trace). The HXR detector observes radiation emitted by REs when
striking first wall or bulk ions and neutrals in plasma, thus both
sources require REs for any signal. The initial rise in HXR correlates
directly with the interferometer signal. There is also good agreement
between the HXR signal the deconfinement of simulated REs. The final
spike in RE fraction lost to the wall is not seen in the HXR
signal. This indicates that future analysis is required to determine
the proportionality constants to the total HXR signal coming from RE
interaction with companion plasma and with wall separately. Lastly,
because the HXR signal does not fully drop to the offset value, it
indicates that there are additional REs remaining after KORC
simulations expect all REs to be deconfined. This is further evidence
that large-angle collisions are generating additional REs when the
induced toroidal electric field is large near the end of the RE beam
deconfinement, and will be discussed more in
Sec.~\ref{sec::conclusion}.

\begin{figure}
\centering
\noindent\includegraphics[width=3.4in]{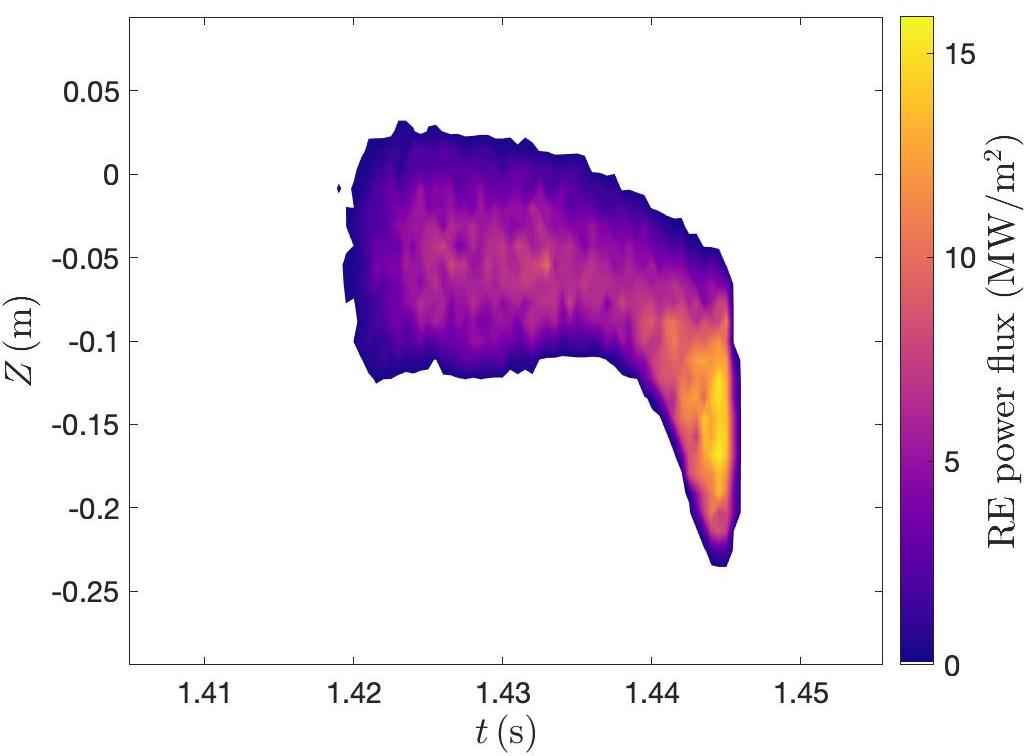}
\caption[]{\label{fig-PowerFlux}Power flux on the inner wall
  calculated for the canonical KORC calculation. The total number of
  experimental REs was estimated and used to scale up the KORC
  calculated REs to arrive at a physically significant value. The
  axisymmetric magnetic field configuration leads to REs impacting the
  inner wall uniformly in the toroidal direction.}
\end{figure}

Because KORC evolves the momentum and location of all simulated
particles, it is possible to calculate the power flux to the wall when
REs are deconfined. As there are of the order $10^{16}$ REs in a
standard RE beam for DIII-D parameters, and only $2.5\times10^4$ REs
simulated with KORC, it is necessary to calculate a scaling factor to
calculate an experimental relevant power flux. We scale the initial
current from DIII-D discharge \#164409 $I_0=2.813\times10^{5}\,{\rm
  A}$, by that calculated using Eq.~(\ref{eq-Ire}) $I_{\rm RE}(0)=
1.15\times10^{-7}$ to get a ratio of $2.45\times10^{12}$. In our
calculations of the power flux, we assume every particle represents a
physical number of particles equal to the calculated ratio. As the
magnetic configuration is axisymmetric, so is the RE deposition on the
inner wall (not shown). The power flux is calculated by binning in the
$Z$ direction in $[-0.3,0.1]\,{\rm m}$, and summing up the particle
energy deposited in a given bin per $0.5\,{\rm ms}$. The calculation
of the power flux varies as the spatial bin width is varied (not
shown); the results presented in Fig.~\ref{fig-PowerFlux} are for $35$
bins of width $1.1\,{\rm cm}$. We note that the deposited power flux
is located below the vertical midplane due to the helicity of the
magnetic field in the present configuration. Future work will include
comparing these deposition power fluxes to infrared camera images.

\section{Conclusions and Discussion}
\label{sec::conclusion}

Simulations performed with the kinetic RE code KORC incorporate
experimentally-reconstructed, time-dependent magnetic and electric
fields, and line integrated electron density data to construct
spatiotemporal models of electron and partially-ionized impurity
transport in the companion plasma. We use KORC to model DIII-D
experiment \#164409 that injects Ne MGI in order to mitigate a
post-disruption, vertically-controlled, RE beam. Comparison of KORC
results and experimental current evolution are performed including
Coulomb collisions with different models of partially-ionized impurity
physics, and it is found that the model presented in
Ref.~\cite{Hesslow17} most closely reproduce the experimental current
evolution.  Comparison of KORC results and experimental current
evolution are performed for different models of initial RE energy and
pitch angle distributions and different spatiotemporal electron and
partially-ionized impurity transport.  The majority of KORC
calculations indicate that while the RE beam current is decreasing,
the RE beam energy increases until confinement degrades. We posit that
collisional pitch angle scattering is primarily responsible for
decreasing the current, while the electric field accelerates REs more
than collisional friction slows them down.  As current dissipates, the
plasma advects toward the inner wall limiter and results in rapid
deconfinement of REs, which we find is the primary dissipation
mechanism, rather than collisions.  Using KORC results on the RE
energy when striking the first wall, we make predictions of the power
flux on the inner wall during RE deconfinement.

The results presented have immediate relevance to ITER and future
reactor level tokamaks.  This work quantifies the efficacy of RE
mitigation via injected impurities, and yields a relative importance
of effects.  The zeroth order effect is the dynamic magnetic field
configuration that determines confinement of REs as summarized by
Fig.~\ref{fig-beamwidthloss} in Sec.~\ref{sec::confinement}.  First
order effects are the inductive toroidal electric field as viewed in
Fig.~\ref{fig-2x1t} of Sec.~\ref{sec::parametric}, the
partially-ionized impurity model as viewed in
Fig.~\ref{fig-boundmodels} of Sec.~\ref{sec::bound}, spatiotemporal
density and partially-ionized impurity transport as viewed in
Fig.~\ref{fig-2x1t} of Sec.~\ref{sec::parametric}, and initial RE
energy and pitch distribution as viewed in Fig.~\ref{fig-2x1t} of
Sec.~\ref{sec::parametric}. Synchrotron and bremsstrahlung radiation
are much smaller effects for the dissipation process in DIII-D
discharge \#164409 and are not shown. A major contribution of the
present study is the identification of the critical role played by
loss of confinement in comparison with the relatively slow collisional
damping.

This work is just the beginning of the necessary KORC modeling efforts
toward understanding important RE spatial effects. Additional preliminary
studies are underway of the spatiotemporal transport of injected
impurities, separately looking at Ar injection, varying the amount of
injected impurities, SPI injection technology, and injection into
different tokamaks, such as JET and KSTAR. We note that the results
presented here are generally in line with recent experiments at JET
exploring high-Z impurity injection for RE mitigation
\cite{Reux20,Lehnen20}.  An important next development for KORC is the
implementation of a large-angle collision operator.  The magnetic to
kinetic energy conversion by large-angle collisions upon the
termination of the RE beam is an ongoing topic of research
\cite{Putvinski97,Loarte11,Hollmann13,MartinSolis14,MartinSolis17}. The
injection of impurities provides additional electrons to knock-on
\cite{Hesslow19}, and the induced toroidal electric field is greatest
when magnetic configuration is rapidly deconfining. The large RE
losses can induce a large toroidal electric field that may increase
the Ohmic current, possibly explaining the results in
Secs.~\ref{sec::confinement},\ref{sec::parametric}, and
\ref{sec::expconnect}.  Because large-angle collisions produces REs at
large pitch angles \cite{Embreus18}, these REs could potentially add a
significant amount of energy to the RE beam without an associated
increase in current. Lastly, tight coupling with an MHD code having
impurity and ablation models is ultimately necessary for robust,
predictive modeling of RE evolution. Such simulations will require
calculation of the self-consistent, induced electric field as the RE
ensemble evolves, and the resulting evolution of the magnetic
configuration evolving with RE current and companion plasma.

\begin{acknowledgments}
  The authors would like to thank J.~Herfindal, E.~Hollmann,
  M.~Lehnen, A.~Lvovskiy, E.~Nardon, C.~Paz-Soldan, and R.~Sweeney for
  their insight into post-disruption, RE experiments, L.~Carbajal,
  M.~Cianciosa, D.~Green, J.~Lore, H.~Lu, and S.~Seal for their help
  with the development of KORC, and J.~Lore and the anonymous
  reviewers for their valuable comments on this manuscript.

  This manuscript has been authored by UT-Battelle, LLC under Contract
  No.~DE-AC05-00OR22725 with the U.S.~Department of Energy. The United
  States Government retains and the publisher, by accepting the
  article for publication, acknowledges that the United States
  Government retains a non-exclusive, paid-up, irrevocable, worldwide
  license to publish or reproduce the published form of this
  manuscript, or allow others to do so, for United States Government
  purposes. The Department of Energy will provide public access to
  these results of federally sponsored research in accordance with the
  DOE Public Access Plan
  (\url{https://www.energy.gov/downloads/doe-public-access-plan}).
  
  This material is based upon work supported by the U.S.~Department of
  Energy, Office of Science, Office of Fusion Energy Sciences, using
  the DIII-D National Fusion Facility, a DOE Office of Science user
  facility, under Award DE-FC02-04ER54698. This research also uses
  resources of the National Energy Research Scientific Computing
  Center (NERSC), a U.S.~Department of Energy Office of Science User
  Facility operated under Contract No.~DE-AC02-05CH11231.

  Data used to generate figures can be obtained in the digital format
  by following the link in Ref.~\cite{data_archive20}.\\
\end{acknowledgments}

\section*{Disclaimer}
  This report was prepared as an account of work sponsored by an
  agency of the United States Government. Neither the United States
  Government nor any agency thereof, nor any of their employees, makes
  any warranty, express or implied, or assumes any legal liability or
  responsibility for the accuracy, completeness, or usefulness of any
  information, apparatus, product, or process disclosed, or represents
  that its use would not infringe privately owned rights. Reference
  herein to any specific commercial product, process, or service by
  trade name, trademark, manufacturer, or otherwise, does not
  necessarily constitute or imply its endorsement, recommendation, or
  favoring by the United States Government or any agency thereof. The
  views and opinions of authors expressed herein do not necessarily
  state or reflect those of the United States Government or any agency
  thereof.

\bibliographystyle{apsrev} \bibliography{bib}

\end{document}